\documentclass[
 reprint,
 amsmath,amssymb,
 aps,nofootinbib, superscriptaddress
]{revtex4-2}

\usepackage{graphicx}
\usepackage{multirow}
\usepackage{threeparttable}         
\usepackage{amssymb}
\usepackage{dcolumn}
\usepackage{bm}
\usepackage{xcolor}
\usepackage[normalem]{ulem}
\usepackage{hyperref}
\graphicspath{{fig}}

\begin{document}
\preprint{APS/123-QED}

\title{Charge imprinting biases topology of correlated insulator in hBN-aligned rhombohedral multilayer graphene}

\author{Lei Qiao,$^{1,*}$ Xin Lu,$^{2,*,\dagger}$ Fu-Chun Zhang,$^{1}$ and Jianpeng Liu$^{2,3,\ddagger}$}
\noaffiliation

\affiliation{Kavli Institute for Theoretical Sciences, University of Chinese Academy of Sciences, Beijing 100190, China}
\affiliation{State Key Laboratory of Quantum Functional Materials, School of Physical Science and Technology, ShanghaiTech Laboratory for Topological Physics, ShanghaiTech University, Shanghai 201210, China}
\affiliation{Liaoning Academy of Materials, Shenyang 110167, China}

\renewcommand{\thefootnote}{\fnsymbol{footnote}}
\setcounter{footnote}{0}
\footnotetext{These authors contributed equally}
\footnotetext{lvxin@shanghaitech.edu.cn}
\footnotetext{liujp@shanghaitech.edu.cn}

\date{\today}

\begin{abstract}
Rhombohedral multilayer graphene aligned with hexagonal boron nitride (RMG-hBN) hosts correlated Chern phases, but the microscopic role of hBN stacking remains unclear, especially when the active carriers are displaced away from the moir\'e interface. Using Hartree-Fock calculations over layer numbers, twist angles, displacement fields, fillings, and hBN alignments, we show that correlated insulators are most robust at small twist angles and intermediate layer number ($N\simeq 6$), where bandwidth suppression is balanced by layer delocalization of the wavefunctions of the active carriers. Under moir\'e-distant conditions at filling $\nu=1$, the topology of the insulating state is strongly biased by charge imprinting: the hBN alignment shapes the occupied valence-band charge texture near the interface via moir\'e potential, which acts through long-range Coulomb interactions as a remote electrostatic template for doped conduction electrons. Depending on the alignment, this template favors either triangular charge localization associated with trivial insulators or honeycomb-like charge networks compatible with Chern insulators. Our results identify valence-band charge textures as a microscopic route by which a remote moir\'e interface controls correlated topology in multilayer graphene.
\end{abstract}

\maketitle

Hexagonal boron nitride-aligned rhombohedral multilayer graphene (RMG-hBN) has emerged as a promising platform for exploring the interplay between strong correlations and topology \cite{lu2024fractional,xie2025tunable,aronson2025displacement,li2025tunable,uzan2025hbn,huo2025does,liu2026odd,dong-R5Gfci-prl-2024,zhou-R5Gfci-prl-2024,dong-R5GAHC-prl-2024,guo2024fractional,kwan2025moire,yu-R5GED-prb-2025,huang-fqahdoublecounting-prb-2024,uchida2026non,kudo2024quantum}. Experiments have revealed a variety of correlated and topological phases tunable by twist angle, layer number, displacement field, and carrier density. Most notably, fractional Chern insulators states \cite{sun2011nearly,regnault2011fractional,neupert2011fractional,sheng2011fractional} have been realized at fractional fillings of moir\'e minibands, experimentally manifested as fractional quantum anomalous Hall effects \cite{cai2023signatures,fqah-nature23,zeng2023thermodynamic,xu2023observation,lu2024fractional,xie2025tunable,aronson2025displacement}. This realizes lattice analogs of fractional quantum Hall states in the absence of an external magnetic field. Since such fractional states often occur concomitantly with flat Chern bands at integer fillings in the same device in RMG-hBN systems \cite{lu2024fractional,xie2025tunable,aronson2025displacement}, understanding how Chern insulators emerge at integer moir\'e filling $\nu$ is a central prerequisite for interpreting and engineering the correlated phase diagrams.

An important but experimentally elusive degree of freedom is the stacking configuration between hBN and the adjacent graphene layer \cite{moon2014electronic,park2023topological,herzog2024moire,jung2014ab,du2020moire}. Depending on whether the boron atom of hBN is aligned with the A or B sublattice of graphene, two inequivalent hBN alignment types arise. Previous theoretical studies have suggested that this stacking configuration can markedly affect the interacting electronic structure and correlated phases in RMG-hBN \cite{kwan2025moire}. However, it is difficult to control during device fabrication and lacks a direct experimental probe. In practice, one may at best realize both configurations within the same device, while leaving their microscopic assignment unresolved \cite{uzan2025hbn, huo2025does}.

This motivates an indirect strategy: inferring the hBN stacking configuration from transport phase diagrams \cite{uzan2025hbn}. Under moir\'e-proximate conditions, where the active carriers reside near the aligned hBN interface under an applied displacement field, Uzan et al. \cite{uzan2025hbn} proposed that the two configurations generate different moir\'e potential strengths through lattice relaxations, leading to distinct phase diagrams consistent with the observations of Aronson \textit{et al.} \cite{aronson2025displacement}. Establishing such an identification scheme for generic RMG-hBN devices, however, requires a systematic theoretical understanding across the experimentally relevant parameter space, including twist angle $\theta$, layer number $N$, displacement field $D$, filling factor $\nu$, and hBN alignment type $\xi$.

A further theoretical puzzle arises under moir\'e-distant conditions, where the displacement field pushes the conduction electrons away from the aligned hBN interface. Calculations indicate that at $\nu=1$, the $\xi=0$ configuration favors Chern-insulating states more robustly than the $\xi=-1$ configuration, even in the absence of lattice-relaxation effects \cite{guo2024fractional,kwan2025moire}. The puzzle thus points to a deeper mechanism: how the moir\'e potential, buried at a remote interface, is felt by active carriers layers away through interactions. Very recently, Herzog-Arbeitman \textit{et al.}~\cite{herzog-arbeitman2026moire,herzog-aps1,herzog-aps2} developed the moir\'e capacitor effect, where moir\'e-modulated valence charges generate an electrostatic potential that stabilizes the $C=1$ parent state and a fractional Chern insulator upon doping. Complementary to this picture, our work reveals how the spatial structure of valence charge imprinting controls topology, identifying this mechanism as the key to resolving the puzzle.

In this work, we address these questions through systematic Hartree-Fock (HF) calculations. We map out correlated phase diagrams as functions of layer number $N$, twist angle $\theta$, filling factor $\nu$, displacement field $D$, and hBN alignment type $\xi$. We find that insulating phases are most abundant at smaller twist angles and for an intermediate number of layers, around $N\simeq 6$, reflecting a competition between enhanced interaction strength and the layer-resolved distribution of wavefunctions with increasing $N$. Most importantly, we identify a real-space charge imprinting mechanism that accounts for the $\xi$ dependence of Chern insulators at $\nu=1$ under moir\'e-distant conditions. We show that hBN alignment reshapes the real-space charge topography of the occupied valence bands. When the conduction electrons reside on the moir\'e-distant side, the valence-band charge profile therefore acts through the long-range interlayer Coulomb interaction as an effective real-space template for conduction electrons, selectively favoring trivial or Chern-insulating states depending on the hBN alignment type. For $\xi=-1$, this imprinted background favors a triangular array of localized charge peaks, promoting a topologically trivial insulating state. By contrast, for $\xi=0$, it favors a connected, honeycomb-like charge network, which stabilizes topologically nontrivial Chern-insulating states. Taken together, our results establish a microscopic theory for understanding how hBN alignment, layer number, twist angle, displacement field, and filling factor conspire to shape correlated topological phases in RMG-hBN systems.

\begin{figure}
    \includegraphics[width=1\linewidth]{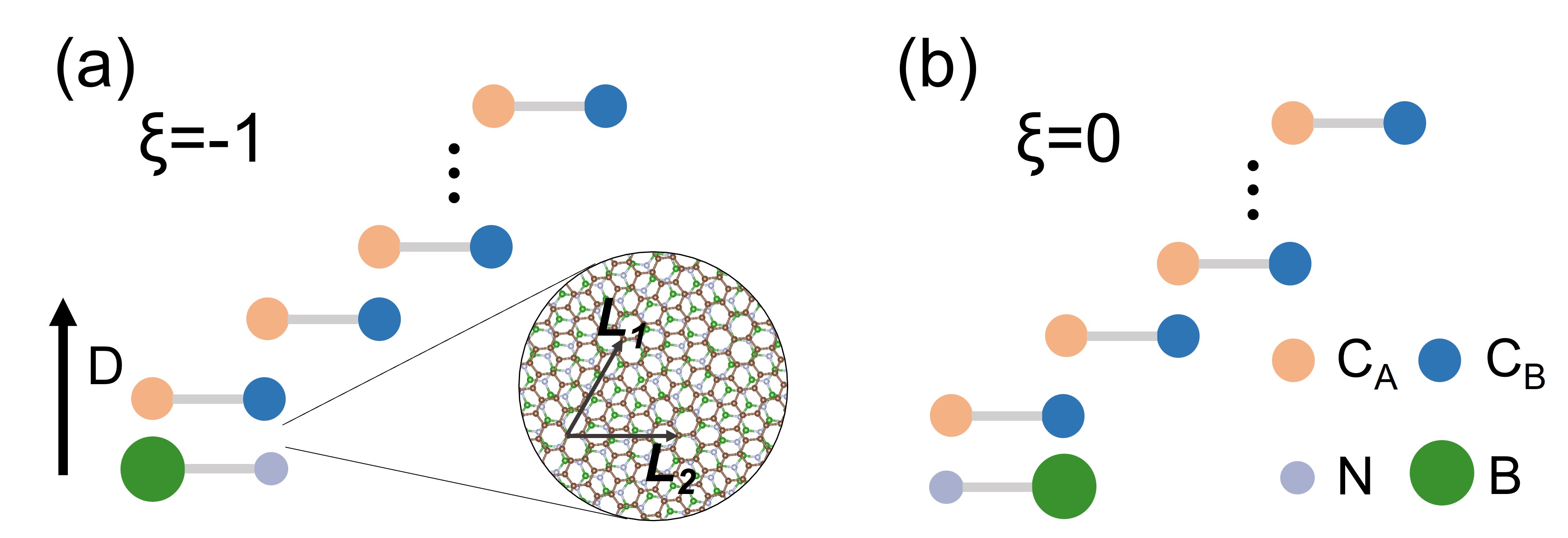}
    \caption{Schematic configurations of RMG twisted relative to a single hBN layer, giving rise to a moir\'e periodicity. Here, $\mathrm{C_A}$ and $\mathrm{C_B}$ denote the A and B sublattices of graphene. In (a), the boron atom of hBN aligns with the A sublattice ($\mathrm{C_A}$), labeled as $\xi = -1$; in (b), the boron atom aligns with the B sublattice ($\mathrm{C_B}$), labeled as $\xi = 0$. Moir\'e interface is zoomed in, shown with moir\'e superlattice vectors, whose lengths range from approximately 10 to 14 nm depending on the twist angle. A positive displacement field is defined as pointing from the moir\'e side toward the non-moir\'e side; that is, conduction band electrons are pushed away from the moir\'e side (``moir\'e distant'').
    }
    \label{fig1}
\end{figure}

We start by introducing the structural configurations of $N$-layer RMG aligned with hBN layer on one side or both sides. When a single hBN interface is present, the alignment is determined by whether the boron atom sits above the A or B sublattice of the adjacent graphene layer. We label these two single-sided configurations by $\xi=-1$ for boron on A and $\xi=0$ for boron on B, as illustrated in Fig.~\ref{fig1}. For devices aligned with hBN on both sides, four double-sided alignments arise, which we summarize in Table~\ref{table1} as $\xi=1,2,3,4$. Throughout this work, we consider the double alignment cases where both hBN layers are twisted by the same angle $\theta$ relative to the graphene; the twist angles we consider are $\theta = 0.201^\circ, 0.449^\circ, 0.77^\circ, 1.08^\circ, 1.36^\circ$, corresponding to moir\'e lengths $L_s \approx 13.5,12.2,10.9,9.5,8.2$\;nm, respectively.

To model the low-energy electronic structure, we employ the continuum Hamiltonian
\begin{equation}\label{eq1}
H_{\text{RMG-hBN}}^{N,\mu} = H_{\text{RMG}}^{N,\mu} + V_{\text{hBN}} \, ,
\end{equation}
where $H_{\mathrm{RMG}}^{N,\mu}$ describes the bare rhombohedral graphene in valley $\mu=\pm$. The moir\'e potential $V_{\mathrm{hBN}}$ is obtained from downfolding the degrees of freedom from hBN \cite{moon2014electronic, guo2024fractional}. Crucially, given one moir\'e interface, the spatial profiles of $V_{\mathrm{hBN}}$ for $\xi=-1$ and $\xi=0$ are related by a $C_{2z}$ rotation, which is the symmetry broken by the hBN interface. To capture $D$-induced charge redistribution, we also determine the layer-resolved electric potentials self-consistently by balancing $D$ with the on-site potential generated by the redistributed interlayer charge density \cite{jang2023chirality, guo2024fractional}. To capture interaction effects, we first apply a perturbative renormalization group approach to Coulomb interactions to work in a low energy window including three conduction and three valence bands per spin per valley \cite{vafek2020renormalization,lu2023synergistic,guo2024fractional}, then perform HF calculations with layer-resolved Coulomb potentials considering the dominant intravalley interactions. A static homogeneous dielectric constant $\epsilon_r=12$ is used throughout.

\begin{table}[h]
    \centering 
    \caption{Six hBN alignment configurations considered. $\xi = -1$ and $0$ denote single-sided alignments, while $\xi = 1, 2, 3, 4$ correspond to double-sided alignments.}
    \label{table1}
    \begin{tabular}{c c c}
        \hline
        hBN-type $\xi$ & Bottom & Top \\
        \hline
        -1 & boron - $\mathrm{C_A}$ & -- \\
        0 & boron - $\mathrm{C_B}$ & -- \\
        1 & boron - $\mathrm{C_B}$ & boron - $\mathrm{C_B}$ \\
        2 & boron - $\mathrm{C_B}$ & boron - $\mathrm{C_A}$ \\
        3 & boron - $\mathrm{C_A}$ & boron - $\mathrm{C_B}$ \\
        4 & boron - $\mathrm{C_A}$ & boron - $\mathrm{C_A}$ \\
        \hline
    \end{tabular}
\end{table}

\begin{figure}
    \includegraphics[width=1\linewidth]{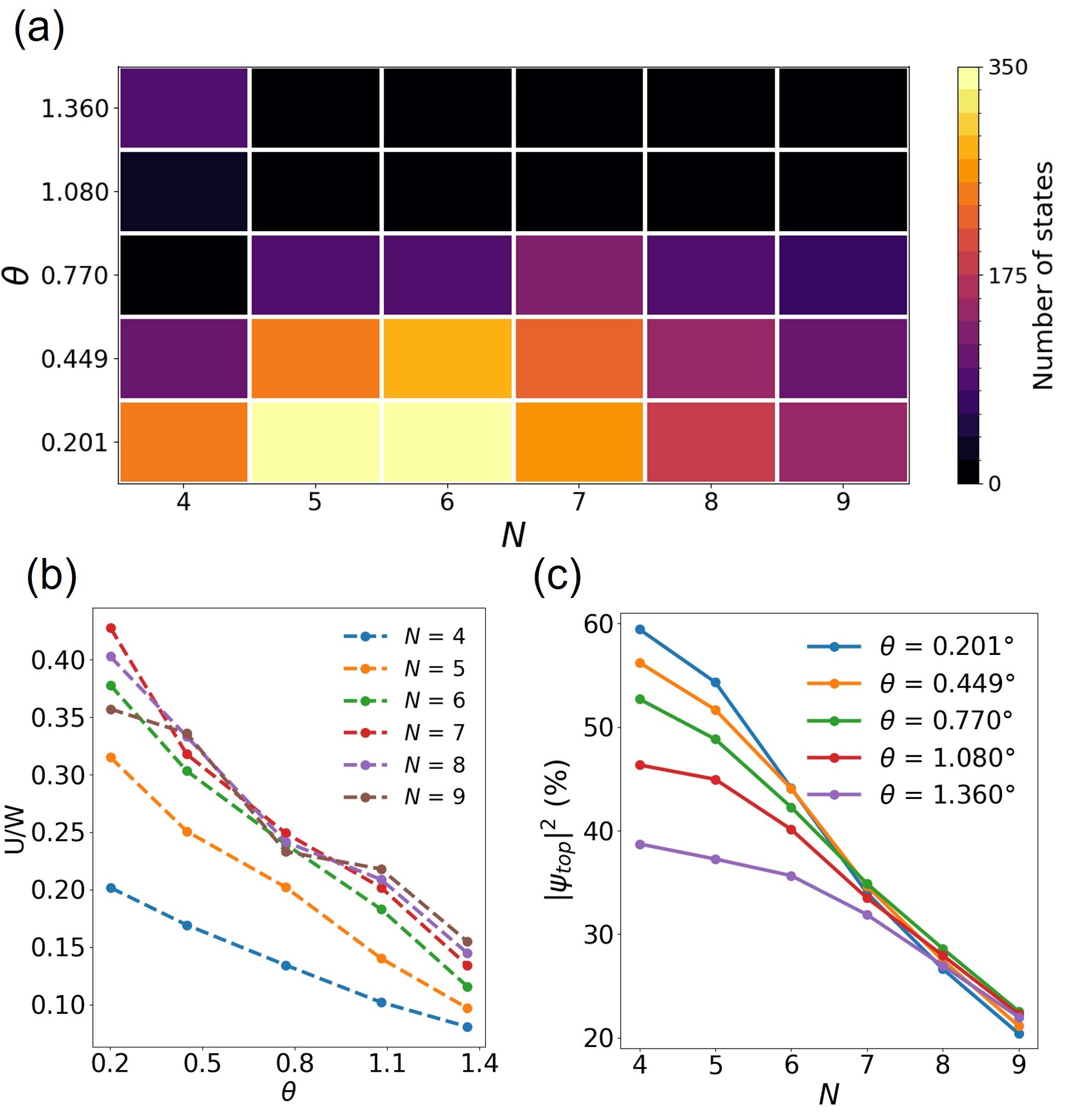}
    \caption{(a) Statistical survey of the number of insulating states across the parameter space spanned by $D$, $\nu$, and $\xi$. (b) Evolution of the interaction strength, defined as the ratio between the typical Coulomb energy $U$ and the bandwidth $W$ as a function of twist angle for different layer number. Here, $U \sim e^2/(4\pi\epsilon_0 \epsilon_r L_s)$ and $W$ is the total bandwidth of the three conduction bands in the low energy window for HF calculations. (c) Evolution of the contribution from the top-layer graphene ($|\psi_{\text{top}}|^2$) to the active frontier conduction bands as a function of layer number for different twist angles. Both (b) and (c) correspond to the case of ($D>0,\nu > 0$) averaged over $D=0.6$-1.1\;V/nm. }
    \label{ins}
\end{figure}

The occurrence of correlated insulating states from spontaneous symmetry breaking is summarized in Fig.~\ref{ins} within the surveyed $(\theta,N)$ parameter space. Each data point aggregates results over the integer fillings $|\nu|=1,2,3$ of the frontier bands, namely the highest valence bands and the lowest conduction bands, all hBN alignment types $\xi=-1,0, 1, 2, 3, 4$, and experimentally relevant displacement fields $0.6\le |D|\le 1.1~{\rm V/nm}$. A clear trend emerges: insulating states occur more frequently at small twist angles for all layer numbers. This behavior reflects the suppression of the kinetic energy scale by the moir\'e length scale. As the twist angle decreases, the moir\'e period increases, the frontier bands become narrower, and interaction-driven gapped states are more readily stabilized, as shown in Fig. \ref{ins}(b) for conduction bands.

Notably, the abundance of insulating states depends non-monotonically on the layer number and reaches a maximum around $N\simeq 6$. This behavior can be understood as the result of two competing tendencies. On the one hand, increasing $N$ generally reduces the bandwidth of the frontier bands, because RMG realizes a higher-order chiral Dirac dispersion in the low-energy two-band description. This bandwidth reduction enhances the relative strength of Coulomb interactions and favors interaction-driven insulating states. On the other hand, within the considered range of displacement fields, the active frontier band wavefunctions become less confined to a single layer (usually one of the lateral surface layers) as $N$ increases, as illustrated in Fig. \ref{ins}(c) for conduction bands under $D>0$. The resulting spread of charge density across multiple layers weakens the effective Coulomb interaction, since intralayer interactions are stronger than interlayer ones. The maximum near $N\simeq 6$ therefore reflects an optimal balance between bandwidth reduction and layer delocalization. Since this mechanism is not tied to a particular hBN alignment type, the same qualitative trend persists when the data are resolved separately for different values of $\xi$, as shown in the alignment-resolved results \cite{supp}. Similar trend also has been observed for $D<0$ or $\nu<0$ \cite{supp}.

A qualitatively different trend appears when we restrict the analysis to Chern-insulating states at $\nu=1$ for $D>0$ in the single-sided hBN-aligned cases, which covers the more than 55\% of the occurrence of Chern insulators in the parameter space. Such states are much more abundant for $\xi=0$ than for $\xi=-1$, and are rarely found in double-sided aligned configurations with $\xi>0$ \cite{supp}. This contrast cannot be accounted for solely by bandwidth reduction or by the overall interaction strength, and instead points to a more specific microscopic mechanism by which the hBN alignment type bias the topology of the insulating state.

\begin{figure}
    \includegraphics[width=1\linewidth]{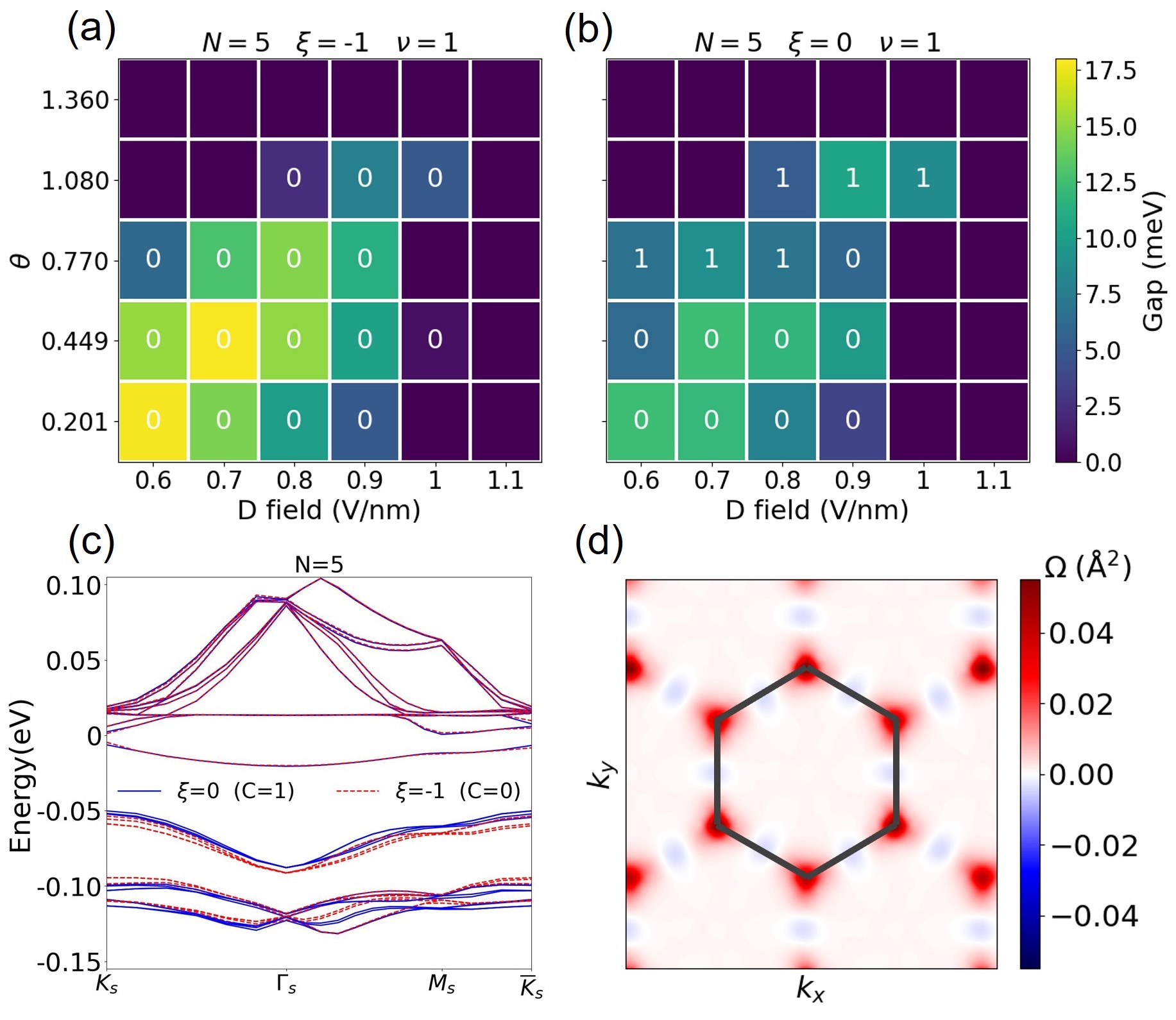}
    \caption{Phase diagrams in the $\theta$--$D$ plane for $N = 5$ moir\'e systems at $\nu = 1$ with configurations (a) $\xi = -1$ and (b) $\xi=0$. The color scale represents the global band gap of the HF band structures with dark purple indicating metallic states. Chern number of the first conduction band is indicated if gapped. (c) HF band structures calculated for $N = 5$ under $D = 0.6$\;V/nm, $\theta = 0.77^\circ$ at $\nu = 1$, including cases with $\xi = -1$ and $\xi = 0$. (d) Berry curvature distributions of the occupied bands for $N = 5$ and $\xi = 0$. The black hexagonal lines delineate the moir\'e Brillouin zone.}
    \label{phase}
\end{figure}

We focus on $N=5$, where the distinction is most pronounced. We map out the HF ground states at $\nu=1$ in the $(\theta,D)$ plane for $N=5$ in Fig.~\ref{phase}(a,b), comparing the two one-sided hBN alignment types $\xi=-1$ and $\xi=0$. Large insulating regions appear at small twist angles, with gaps reaching up to $18~{\rm meV}$. Upon increasing $\theta$, the moir\'e period decreases and the kinetic energy scale grows, reducing the effective interaction strength relative to the bandwidth; consequently, the insulating regions shrink and eventually disappear. Although the two alignment types produce phase diagrams of comparable overall extent for gapped states, their topology differs qualitatively: for $\xi=0$, a finite window of Chern-insulating states emerges, whereas for $\xi=-1$ the insulating states are predominantly topologically trivial.

This contrast is striking because the HF band dispersions themselves can be nearly indistinguishable. Fig.~\ref{phase}(c) shows representative HF band structures for $N=5$, $\theta=0.77^\circ$ under $D=0.6$\;\rm V/nm at $\nu=1$. The band dispersions for $\xi=-1$ (red dashed lines) and $\xi=0$ (solid blue lines) are almost identical, yet only the latter exhibits an isolated conduction band carrying nonzero total Berry curvature peaking at the two moir\'e $K$ points [Fig.~\ref{phase}(d)]. This provides a striking example in which the band topology is largely decoupled from the energy dispersion. The distinction must therefore originate from wavefunction properties that are invisible in the energy spectrum but nevertheless influence the correlated ground-state topology.

\begin{figure}
    \includegraphics[width=0.95\linewidth]{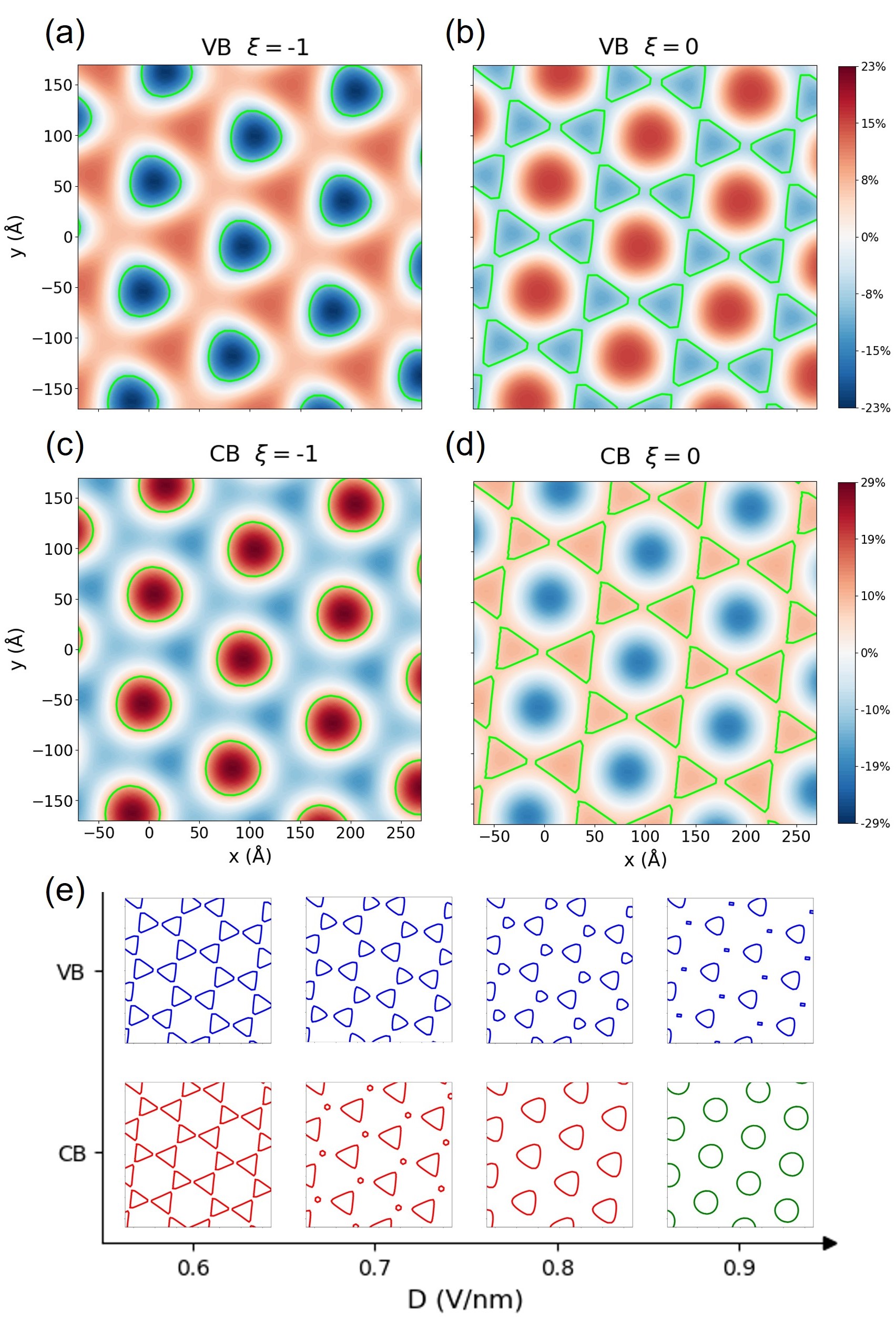}
    \caption{Real-space charge density profile of (a, b) the valence bands and (c, d) the first conduction band for $N = 5$, $\nu = 1$, $\theta = 0.77^\circ$. Panels (a, c) and (b, d) represent the charge density with $\xi = -1$ ($C = 0$) and $\xi = 0$ ($C = 1$), respectively, where $D=0.6$ V/nm. The results for $D=0.7$-$0.9$\;V/nm are provided in the Supplemental Material~\cite{supp}. The color code represents charge density fluctuation defined as $\delta\rho = (\rho - \bar{\rho}) / \bar{\rho}_{0} \times 100\%$, where $\rho$ is the electron density normalized to the number of occupied bands, and $\bar{\rho}_{0}$ is the average density of a single electron per moir\'e unit-cell. For illustrative purpose, the lines indicate contours at $44$\% ($34$\%) of the maximum of $|\delta\rho|$, marking the low-density pockets of valence charges (high-density peaks of excess conduction charges). Panel (e) shows the evolution of the low(high)-density regions of valence (conduction) electrons for $\xi=0$ as a function of $D$. For conduction bands, the high-density region of the $C=1$ state is encircled by red lines, while the $C=0$ by green lines; low-density regions for valence bands are encircled by blue lines.}
    \label{rho}
\end{figure}

We identify the relevant topological properties with the real-space charge distributions. Topologically distinct insulating states can possess qualitatively different real-space charge patterns \cite{zeng2024sublattice}. This distinction is particularly pronounced in the moir\'e-distant regime considered here. The moir\'e potential generated by the aligned hBN substrate decays rapidly through the graphene stack, whereas the Coulomb interaction remains long-ranged. Consequently, the occupied valence-band charge density, that is concentrated predominantly on the bottom, hBN-proximal layers, is strongly modulated by the hBN alignment type. Even when a positive displacement field pushes the doped conduction electrons away from the aligned interface, they remain sensitive to the electrostatic landscape imprinted by the filled valence bands. 

This effect is illustrated for $N=5$, $\theta=0.77^\circ$, and $D=0.6$~V/nm. For $\xi=-1$, the excess valence charge forms an interconnected ring-like network, leaving isolated low-density pockets [Fig.~\ref{rho}(a)]. The doped conduction electrons avoid the high-density valence background and localize in these pockets, producing a triangular charge pattern [Fig.~\ref{rho}(c)]. By contrast, for $\xi=0$, the valence charge profile leaves a connected honeycomb-like network of low-density regions, with six local minima per moir\'e unit cell [Fig.~\ref{rho}(b)]. The doped electrons can therefore occupy an effective honeycomb lattice [Fig.~\ref{rho}(d)] to minimize interlayer Coulomb energy. Importantly, stabilization through the formation of an interlocking charge pattern mediated by long-range Coulomb interactions is not unique to RMG-hBN systems. More generally, a correlated insulator may be stabilized by a periodic potential that favors an interlocking pattern with its charge density, whether the potential originates from a different substrate or charge ordering in another correlated system, as suggested by recent works \cite{lu2023synergistic,zhou2021bilayer}.

The two charge textures have distinct topological implications. If the nontrivial topology is tied to an emergent sublattice degree of freedom visible in the real-space charge texture~\cite{zeng2024sublattice}, namely to the winding of an effective sublattice pseudospin across the moir\'e Brillouin zone, then a triangular charge pattern, which lacks such a sublattice structure, is naturally associated with a topologically trivial state. Conversely, a honeycomb-like charge network provides the minimal real-space structure compatible with a nontrivial sublattice pseudospin texture, although it is not by itself sufficient to guarantee a nonzero Chern number. Consistent with this interpretation, in all cases we have examined, the isolated flat band of the Chern insulator exhibits a connected charge pattern, whereas that of the trivial insulator exhibits an isolated charge pattern, also observed in Refs.~\cite{zeng2024sublattice,guo2025correlation,kwan2025moire}. Thus, in the $\xi=0$ configuration, the valence-band electrostatic imprint favors a connected honeycomb-like conduction charge texture and thereby biases the system toward a Chern insulator. In contrast, the triangular localization tendency induced by the $\xi=-1$ valence background favors a topologically trivial insulator. 

Further support for this charge imprinting mechanism comes from the evolution of the valence- and conduction-band charge profiles with increasing displacement field in the $\xi=0$ configuration [Fig.~\ref{rho}(e)]. As $D$ is increased from $0.6$ to $0.9$\;V/nm, the HF ground state evolves from a $C=1$ to a $C=0$ gapped state. Over the same range, the honeycomb pattern formed by the low-density regions of the valence charge develops an increasing sublattice asymmetry. The high-density peaks of the first conduction band show the same trend, but more prominently: near the phase transition at $D=0.8$~V/nm, the excess charge remains connected \cite{supp}, yet its maxima become strongly polarized toward one sublattice, approaching a triangular charge texture. In fact, the triangular texture can be viewed as the fully sublattice-polarized limit of the honeycomb texture. An analogous evolution is also observed in momentum space, where the Berry curvature progressively concentrates near one of the two moir\'e $K$ points as the transition is approached~\cite{supp}, showing the duality of topology between real-space charge density and reciprocal space Berry curvature distribution. Upon further increasing $D$ to $0.9$~V/nm, the trivial insulator becomes the HF ground state once the sublattice asymmetry of the valence charge profile exceeds a critical strength. We emphasize that signatures of this valence-band imprinting are already present at the non-interacting level. Hartree and Fock interactions enhance the charge modulation of the conduction electrons, but do not generate the underlying alignment-dependent texture from scratch, as demonstrated by a systematic comparison of non-interacting, Hartree-only, and full HF charge densities~\cite{supp}.

The importance of charge imprinting can be quantitatively evaluated by computing the interband Coulomb-energy contribution to the competition between the $C=1$ and $C=0$ states. When the valence-band charge profile favors a honeycomb conduction texture, as in the $\xi=0$ case, a competing triangular $C=0$ state incurs an additional Coulomb cost because its preferred charge distribution is incompatible with the imprinted electrostatic landscape. Conversely, when the valence background favors a triangular conduction texture, as in the $\xi=-1$ case, a competing honeycomb-like $C=1$ state is penalized.

This point is most clearly seen for $N=7$, $\theta=1.08^\circ$, $\nu=1$, and $D=0.68$~V/nm. For $\xi=-1$, the HF ground state is a $C=0$ insulator, while a $C=1$ state appears as a nearby metastable solution. For $\xi=0$, the ordering is reversed, with a $C=1$ HF ground state competing with a metastable $C=0$ state. In both cases, the HF band structures of the two competing solutions are nearly identical \cite{supp}, and their kinetic-energy difference is negligible compared with the interaction-energy difference. For $\xi=-1$, where the $C=0$ state is the HF ground state, the difference in the interband Hartree contribution is $0.18$~meV per moir\'e unit cell, whereas the total energy difference is only $0.03$~meV per moir\'e unit cell. Thus, omitting the interband Hartree term would reverse the energetic ordering and make the $C=1$ state the ground state, demonstrating the dominant role of charge imprinting in this case. For $\xi=0$, where the $C=1$ state is the HF ground state, the total energy difference is approximately $0.35$~meV per moir\'e unit cell. About $20\%$ of this energy difference arises from the interband Coulomb contribution, while about $70\%$ comes from intraband Coulomb interactions within the conduction bands. In such case, charge imprinting is a non-negligible contribution to the stabilization of the Chern insulator, although the dominant energetic contribution in this case comes from intraband interactions. 

We therefore do not regard charge imprinting as the sole factor controlling the competition between the $C=0$ and $C=1$ phases, although it is a crucial factor that plays a dominating role in many cases. Rather, it should be viewed as an alignment-dependent energetic bias toward a particular real-space charge texture. This picture is consistent with previous theoretical~\cite{kwan2025moire, nashabeh2026lattice} and experimental~\cite{huo2025does} results, where $C=1$ states can appear for both $\xi$. In the same spirit, we also find that a $C=1$ state can become the ground state even for $\xi=-1$, for example for $N=7$, $\theta=1.08^\circ$, $\nu=1$, and $D=0.7$~V/nm~\cite{supp}. Such reversed situations occur more frequently in thicker stacks. This trend can be understood as a weakening of the valence-band imprinting effect:  within the relevant range of displacement fields, increasing the layer number on the one hand makes the valence-band wavefunctions more delocalized across the stack, thus are less affected by the interface moir\'e potential; and on the other hand increases the spatial separation between the valence charge near the hBN interface and the doped conduction charge on the opposite side. Both effects reduce the electrostatic bias imposed by the occupied valence band charge textures, allowing other energetic contributions, such as intraband interactions, to play a more prominent role in selecting the ground-state topology.

Beyond the $\nu=1$ case emphasized above, we also find a $C=1$ insulator at $\nu=2$, where two isolated conduction bands carry Chern numbers $0$ and $1$, respectively~\cite{supp}, also observed in previous theoretical and experimental results~\cite{kudo2024quantum,liu2026odd,zhang2025entangled, uchida2026non}. The broader phase diagram contains several additional competing gapped states, including spin-polarized, valley-polarized, spin-valley-polarized, and intervalley-coherent states~\cite{kudo2024quantum,uzan2025hbn,liu2026odd}. A full classification of these phases lies beyond the scope of this work. We also leave a comprehensive study of moir\'e-proximate regime and double-sided alignments with $\xi>0$ for future work, since in these cases the active carriers can be driven close to at least one moir\'e interface, where lattice relaxation and other microscopic interface effects are expected to become more important~\cite{uzan2025hbn, nashabeh2026lattice}.

In summary, we have performed a systematic HF study of RMG-hBN as a function of layer number, twist angle, displacement field, filling, and hBN alignment. First, we show that correlated insulating states are favored by small twist angles but are most abundant at intermediate layer number, reflecting the balance between bandwidth suppression and layer delocalization of the wavefunctions of active carriers. Second, under moir\'e-distant conditions, hBN alignment biases the topology of the $\nu=1$ insulator through charge imprinting: the occupied valence bands near the hBN interface form alignment-dependent charge textures that act, through long-range Coulomb interactions, as an electrostatic template for doped conduction electrons on the opposite side of the stack. This template favors either triangular charge localization, which is biased toward a trivial insulator, or a connected honeycomb-like texture compatible with an emergent sublattice pseudospin and a Chern insulator. Because charge imprinting supplies an energetic bias rather than a deterministic topological criterion, competing intraband interactions and layer delocalization can reverse the energetic ordering, especially in thicker stacks. These results provide a microscopic basis for interpreting stacking-dependent phase diagrams and for engineering Chern phases through remote real-space charge textures in multilayer graphene moir\'e systems.

\section*{Acknowledgements} 
 Xin Lu thanks the financial support from the Quantum Science and Technology-National Science and Technology Major Project (grant no. 2025ZD0300500) and the National Natural Science Foundation of China (grant no. 12404221). Jianpeng Liu acknowledges support from the National Key Research and Development Program of China (grant no. 2024YFA1410400), the Strategic Priority Research Program of Chinese Academy of Sciences (grant no. XDB1710000), the National Natural Science Foundation of China (grant no. 12550403) and Shanghai Science and Technology Innovation Action Plan (grant no. 24LZ1401100). Lei Qiao acknowledges support from China Postdoctoral Science Foundation (2025M773388). Fuchun Zhang acknowledges support from the National Natural Science Foundation of China (grant no. 12574150). Xin Lu also acknowledges helpful discussions with Jonah Herzog-Arbeitman.

\bibliographystyle{apsrev4-2}
\bibliography{ref}

\end{document}

\graphicspath{{fig}}

\preprint{APS/123-QED}

\title{Supplemental Materials for ``Charge imprinting biases topology of correlated insulator in hBN-aligned rhombohedral multilayer graphene''}

\author{Lei Qiao,$^{1,*}$ Xin Lu,$^{2,*,\dagger}$ Fu-Chun Zhang,$^{1}$ and Jianpeng Liu$^{2,3,\ddagger}$}
\noaffiliation

\affiliation{Kavli Institute for Theoretical Sciences, University of Chinese Academy of Sciences, Beijing 100190, China}
\affiliation{State Key Laboratory of Quantum Functional Materials, School of Physical Science and Technology, ShanghaiTech Laboratory for Topological Physics, ShanghaiTech University, Shanghai 201210, China}
\affiliation{Liaoning Academy of Materials, Shenyang 110167, China}

\maketitle

\vspace{-0.8cm}
\begin{center}
\footnotesize
$^{*}$ These authors contributed equally\\
$^{\dagger}$ lvxin@shanghaitech.edu.cn\\
$^{\ddagger}$ liujp@shanghaitech.edu.cn
\end{center}
\vspace{0.2cm}

\renewcommand{\theequation}{S\arabic{equation}}
\renewcommand{\figurename}{Supplementary Figure}

\def\Red#1{\textcolor{red}{#1}}
\def\Blue#1{\textcolor{blue}{#1}}
\def\Green#1{\textcolor{green}{#1}}

% \appendix
\tableofcontents

\section{Non-interacting continuum model for multi-layer graphene-hBN moir\'e system}\label{continum model}

In the RMG-hBN system, hBN acts as a wide band gap semiconductor and does not directly contribute low-energy electronic states. Therefore, its Hamiltonian can be separated into two parts: the graphene component $H^{N,\nu}_{RMG}$ and the hBN-induced potential $V_{hBN}$. We begin by deriving the $H^{N,\nu}_{RMG}$ part from a tight-binding model. The lattice vectors of graphene are defined as $\textbf{a}_1=a_0(1,0)$ and $\textbf{a}_2=a_0(1/2,\sqrt{3}/2)$, with the lattice constant $a_0=2.46$ $\mathring{\mathrm{A}}$. The coordinates of the A and B sublattices are denoted as $\tau_{1,A}=(0,a_0/\sqrt{3})$ and $\tau_{1,B}=(0,0)$ within the first layer $l=1$, while the corresponding reciprocal lattice vectors are given by $\textbf{b}_1=4 \pi / \sqrt{3}a_0(\sqrt{3}/2,-1/2)$ and $\textbf{b}_2=4 \pi / \sqrt{3}a_0(0,1)$, respectively. For the rhombohedral stacked graphene considered in our study, the adjacent layers adopt Bernal stacking, which involves $(0,-a_0\sqrt{3})$ shift, meaning that the B sublattice of the lower layer overlaps with the A sublattice of the upper layer from a top view. Consequently, in the $l$-th layer, the positions of the A and B sublattices are $\tau_{l,A}=(2-l) \times(0,a_0/\sqrt{3})$ and $\tau_{l,B}=(1-l) \times(0,a_0/\sqrt{3})$, respectively. Based on the structural, the Slater–Koster tight-binding Hamiltonian can be written as:
\begin{equation}
    H_{RMG}=\sum_{i,j,l_1,l_2,\alpha,\beta} -t (\mathbf{R}_i+\tau_{l_1,\alpha} + l_1d_0\mathbf{e}_z
    -\mathbf{R}_j-\tau_{l_2,\beta} - l_2d_0\mathbf{e}_z)
    \hat{c}^\dagger_{l_1,\alpha}(\mathbf{R}_i)\hat{c_{l_2,\beta}(\mathbf{R}_j)}
\end{equation}
where $(i,j),(l_1,l_2)$ and $(\alpha,\beta)$ are the unit-cell, layer and sublattice indices, respectively. We adopt the graphene interlayer distance $d_0=3.35$ $\mathring{\mathrm{A}}$, and $\mathbf{e}_z$ is the unit vector perpendicular to the graphene plane. Here we only consider the hopping between the neighbor graphene sheets, and the hopping parameters are selected to be common Slater-Koster-type function for any combinations of atomic species:
\begin{subequations}\label{eq:S2}
\begin{align}
-t(\mathbf{R}) &= V_{pp\pi} \left[ 1 - \left( \frac{\mathbf{R}\cdot \mathbf{e}_z}{|\mathbf{R}|} \right)^2 \right]
   + V_{pp\sigma} \left( \frac{\mathbf{R}\cdot \mathbf{e}_z}{|\mathbf{R}|} \right)^2 , \label{eq:S2a} \\[6pt]
V_{pp\pi} &= V_{pp\pi}^0 \exp\!\left( -\frac{|\mathbf{R}| - a_0/\sqrt{3}}{r_0} \right) , \label{eq:S2b} \\[6pt]
V_{pp\sigma} &= V_{pp\sigma}^0 \exp\!\left( -\frac{|\mathbf{R}| - d_0}{r_0} \right) . \label{eq:S2c}
\end{align}
\end{subequations}
we take the parameters $V_{pp\pi}^0=-2.7$ eV, $V_{pp\sigma}^0=0.48$ eV and $r_0=0.184a_0$ from prior work\cite{moon2014electronic}. Next, we perform the following Fourier transform of the tight-binding Hamiltonian within the plane of the graphene layers:
\begin{subequations}
\begin{align}
    \hat{c}_{l,\alpha}(\mathbf{R})&=\frac{1}{\sqrt{N_c}}\sum_\mathbf{k} e^{i\mathbf{k}\cdot(\mathbf{R}+\tau_{l,\alpha})}\hat{c}_{l,\alpha}(\mathbf{k})\\
    \hat{c}_{l,\alpha}(\mathbf{k})&=\frac{1}{\sqrt{N_c}}\sum_\mathbf{k} e^{-i\mathbf{k}\cdot(\mathbf{R}+\tau_{l,\alpha})}\hat{c}_{l,\alpha}(\mathbf{R})
\end{align}
\end{subequations}
where $N_c$ is the number of unit cells.

Expanding the above tight-binding Hamiltonian $H_{RMG}$ around the Dirac points $\mathbf{K/K}^\prime$ (also noted as $\mathbf{K^+/K^-}$, respectively, with $\mathbf{K}^\mu =-\mu (4\pi/3a_0,0)$) and $\mu=\pm1$), the following $\mathbf{k\cdot p}$ Hamiltonian of $N$ layer RMG-hBN system for valley $\mu$ reads:
\begin{equation}
H^{N,\mu}_{\text{RMG}} =
\begin{pmatrix}
h^{1,\mu}_{\text{intra}} & \left(h^{\mu}_{\text{inter}}\right)^\dagger & 0 & 0 & 0 \\
h^{\mu}_{\text{inter}} & h^{2,\mu}_{\text{intra}} & \left(h^{\mu}_{\text{inter}}\right)^\dagger & 0 & 0\\
0 & h^{\mu}_{\text{inter}} & h^{3,\mu}_{\text{intra}} & \ddots & \vdots\\
0 & 0 & \ddots & \ddots & \left(h^{\mu}_{\text{inter}}\right)^\dagger \\
0 & 0 & \cdots & h^{\mu}_{\text{inter}} & h^{N,\mu}_{\text{intra}}
\end{pmatrix}
\end{equation}
Here, the intralayer Hamiltonian $h_{intra}^{l,\mu}$ is given by the $k\cdot p$ model for monolayer graphene:
\begin{equation}
    h_{intra}^{l,\mu}= -\hbar v_F \mathbf{k} \cdot \sigma_\mu
\end{equation}
where $\hbar v_F=5.253$ eV$\cdot \mathring{\mathrm{A}}$ is the non-interacting Fermi velocity of Dirac fermions extracted from the Slater-Koster tight-binding model, $\mathbf{k}$ is the wave vector expanded around the Dirac point in valley $\mu$, and $\sigma_\mu=(\mu\sigma_x,\sigma_y)$ is the
Pauli matrix defined in sublattice space. The interlayer coupling $h_{inter}^{\mu}$ is given by:
\begin{equation}
    h^\mu_{\text{inter}}=
\begin{pmatrix}
    \hbar v_{\perp}(\mu k_x +ik_y) & t_\perp \\
    \hbar v_\perp (\mu kx_x - i k_y) & \hbar v_\perp (\mu k_x + ik_y)\\
\end{pmatrix}
\end{equation}

Next, we turn to the hBN-induced potential term $V_{\mathrm{hBN}}$. The hBN layer has a honeycomb lattice similar to that of graphene, but with a slightly larger lattice constant. We define the stacking such that N atoms occupy the A sites and B atoms occupy the B sites. Starting from the configuration without relative rotation, the A\text{--}B bonds in graphene and hBN are parallel. In this case, the A site of graphene can align either with the A site or the B site of hBN, corresponding to $\xi=0$ and $-1$, respectively. After rotating hBN with respect to graphene by an angle $\theta$, the lattice vectors of hBN become:
\begin{equation}
    \tilde{\mathbf{a}}_{1,2}=M R(\theta) \mathbf{a}_{1,2}
\end{equation}
where $M=a_{hBN}/a_0$ and $R(\theta)$ is the rotation matrix. Therefore, the mismatch between the two lattices generates
moir\'e pattern with the moir\'e lattice vectors $\mathbf{L}_{1,2}$ and reciprocal lattice vectors $\mathbf{G}_{1,2}$:
\begin{subequations}
\begin{align}
    \mathbf{L}_{1,2}&=(1-R(-\theta)M^{-1})\mathbf{a}_{1,2}\\
    \mathbf{G}_{1,2}&=(1-M^{-1}R(\theta))\mathbf{b}_{1,2}
\end{align}
\end{subequations}

We can write the effective Hamiltonian in the following form:
\begin{equation}
    H_{\mathrm{RMG-hBN}}=\begin{pmatrix} H_{\mathrm{RMG}} & U^{\dagger} \\ U & H_{\mathrm{hBN}}\end{pmatrix} \\
\end{equation}
where $U$ is the interlayer coupling term between graphene and hBN. Since the wide-gap insulator hBN does not contribute to the low-energy electronic states, we can safely neglect the dispersion by dropping the $\mathbf{k}$ dependence in $H_{\mathrm{hBN}}$:
\begin{equation}
    H_{\mathrm{hBN}}=\begin{pmatrix}
       V_{\mathrm{N}} & 0 \\ 0 & V_{\mathrm{B}}
    \end{pmatrix}
\end{equation}

For $\theta=0.201^\circ,0.449^\circ,0.770^\circ,1.080^\circ$ and $1.360^\circ$, the corresponding moir\'e length $L_s=|\mathbf{L}_{1,2}|$ are 13.5, 12.2, 10.9, 9.5 and 8.2 nm. 
When $\theta$ is small, the local structure in the moir\'e supercell can be regarded as a bilayer with a relative displacement $\delta$ without rotation, $\delta(\mathbf{r})=(1-R^{-1}(\theta)M^{-1})\mathbf{r}$, which varies slowly with position $\mathbf{r}$. Therefore, $U$ can be written as\cite{moon2014electronic}:
\begin{equation}
    U=u_0\left[\begin{pmatrix}1 & 1 \\1 & 1\end{pmatrix} + 
    \begin{pmatrix}1 & \omega^{-\mu} \\ \omega^\mu & 1\end{pmatrix} e^{i\mu \mathbf{G}_1 \cdot \mathbf{r}}
    + \begin{pmatrix}1 & \omega^{\mu} \\ \omega^{-\mu} & 1\end{pmatrix} e^{i\mu (\mathbf{G}_1 + \mathbf{G}_2) \cdot \mathbf{r}}
    \right]
\end{equation}
where $\omega = e^{2\pi i /3}$ and $u_0=0.152$ eV.

Below, we transform $H_{\mathrm{hBN}}$ into an effective moir\'e superlattice potential that acts directly on the adjacent graphene layer:
\begin{equation}
    V_{\text{hBN}}=-U^{\dagger}H_{\mathrm{hBN}}U=V_{\text{eff}}(\mathbf{r})+M_\mathrm{eff}(\mathbf{r})\sigma_z + e v_F\mathbf{A}_{\mathrm{eff}}(\mathbf{r}) \cdot \sigma_\mu
\end{equation}
where the scalar potential $V_{\mathrm{eff}}$, the vector potential $\mathbf{A}_{\mathrm{eff}}$, and the the Dirac mass term $M_{\mathrm{eff}}$ reads:
\begin{subequations}
\begin{align}
V_{\mathrm{eff}}(\mathbf{r}) &= V_0 - V_1 \sum_{j=1}^3 \cos \alpha_j(\mathbf{r}) \\
M_{\mathrm{eff}}(\mathbf{r}) &= \sqrt{3} V_1 \sum_{j=1}^3 \sin \alpha_j(\mathbf{r}) \\
e v_F \mathbf{A}_{\mathrm{eff}}(\mathbf{r}) &= 2 \mu V_1 \sum_{j=1}^3 
\begin{pmatrix}
\cos\!\left[ \tfrac{2\pi}{3}(j+1) \right] \\
\sin\!\left[ \tfrac{2\pi}{3}(j+1) \right]
\end{pmatrix}
\cos \alpha_j(\mathbf{r}) \\
\alpha_j(\mathbf{r}) &= \mathbf{G}_j \cdot \mathbf{r} + \psi + \tfrac{2\pi}{3}, 
\qquad \text{with } \mathbf{G}_3 = -\mathbf{G}_1 - \mathbf{G}_2
\end{align}
\end{subequations}
where $V_0=0.0289$ eV, $V_1=0.0210$ eV, and $\psi$ is 2.38 (-0.29) for $\xi=-1(0)$. Finally, the continuum model Hamiltonian for RMG-hBN system with $\xi\leq 0$ can be written as:
\begin{equation}
H^{N,\mu}_{\text{RMG-hBN}} =
\begin{pmatrix}
h^{1,\mu}_{\text{intra}}+V_{\mathrm{hBN}} & \left(h^{\mu}_{\text{inter}}\right)^\dagger & 0 & 0 & 0 \\
h^{\mu}_{\text{inter}} & h^{2,\mu}_{\text{intra}} & \left(h^{\mu}_{\text{inter}}\right)^\dagger & 0 & 0\\
0 & h^{\mu}_{\text{inter}} & h^{3,\mu}_{\text{intra}} & \ddots & \vdots\\
0 & 0 & \ddots & \ddots & \left(h^{\mu}_{\text{inter}}\right)^\dagger \\
0 & 0 & \cdots & h^{\mu}_{\text{inter}} & h^{N,\mu}_{\text{intra}}
\end{pmatrix}
\label{finalh}
\end{equation}
For those configurations with $\xi > 0$, an additional potential $V_{\mathrm{hBN}}$ should be applied to $h^{N,\mu}_{\text{intra}}$.

\section{Hartree-Fock approximations to the electron-electron interactions}\label{HF}

We now turn to the electron-electron interactions that impact the low-energy behavior, which can be written as:
\begin{equation}
    \hat{V}_{ee}=\frac{1}{2} \int d^2rd^2r'\sum_{\sigma.\sigma'} 
    \hat{\psi}_\sigma^\dagger(\mathbf{r}) \hat{\psi}_{\sigma'}^\dagger V_c(|\mathbf{r}-\mathbf{r'}|)\hat{\psi}_{\sigma'}(\mathbf{r}')\hat{\psi}_{\sigma}(\mathbf{r})
\end{equation}
where $\hat{\psi}_{\sigma}(\mathbf{r})$ is the electron annihilation operator at position $\mathbf{r}$ and spin $\sigma$. This interaction can be rewritten as:
\begin{equation}
    \hat{V}_{ee}=\frac{1}{2}\sum_{ii'jj'}\sum_{ll'mm'}\sum_{\alpha\alpha'\beta\beta'}\sum_{\sigma\sigma'}\hat{c}_{i,\sigma l \alpha}^\dagger\hat{c}_{i',\sigma' l' \alpha'}^\dagger V_{ij,i'j'}^{\alpha\beta l m \sigma,\alpha'\beta' l' m' \sigma'}\hat{c}_{j',\sigma' m' \beta'}\hat{c}_{j,\sigma m \beta}
\end{equation}
where 
\begin{equation}
\begin{aligned}
V^{\alpha \beta l m \sigma, \alpha' \beta' l' m' \sigma'}_{i j, i' j'} 
&= \int d^2 r \, d^2 r' \, V_c(|\mathbf{r} - \mathbf{r}'|) \,
\phi^*_{l,\alpha}(\mathbf{r} - \mathbf{R}_i - \boldsymbol{\tau}_{l,\alpha}) \,
\phi_{m,\beta}(\mathbf{r} - \mathbf{R}_j - \boldsymbol{\tau}_{m,\beta}) \\
&\quad \times \phi^*_{l',\alpha'}(\mathbf{r} - \mathbf{R}'_i - \boldsymbol{\tau}_{l',\alpha'}) \,
\phi_{m',\beta'}(\mathbf{r} - \mathbf{R}'_j - \boldsymbol{\tau}_{m',\beta'}) \,
\chi^\dagger_\sigma \chi^\dagger_{\sigma'} \chi_{\sigma'} \chi_\sigma
\end{aligned}
\end{equation}
where $i, \alpha$ and $\sigma$ refer to Bravais lattice vectors, layer/sublattice index, and spin index. $\phi$ is Wannier function and $\chi$ is the two-component spinor wave function. Due to the narrow bandwidth, we can further assume the dominant part is electron-electron interaction, hence the above interaction can be simplified to:
\begin{equation}
\begin{aligned}
\hat{V}_{ee} &= \tfrac{1}{2} \sum_{ii'} \sum_{\alpha \alpha' l l'} \sum_{\sigma \sigma'} \hat{c}^\dagger_{i,\sigma l \alpha} \hat{c}^\dagger_{i',\sigma' l' \alpha'} 
V_{i\sigma l \alpha,\, i' \sigma' l' \alpha'} \hat{c}_{i',\sigma' l' \alpha'}  \hat{c}_{i,\sigma l \alpha} \\ &\approx \tfrac{1}{2} \sum_{i l \alpha \neq i' l' \alpha'} \sum_{\sigma \sigma'} \hat{c}^\dagger_{i,\sigma l \alpha} \hat{c}^\dagger_{i',\sigma' l' \alpha'} V_{i l \alpha,\, i' l' \alpha'} \, \hat{c}_{i',\sigma' l' \alpha'} 
\hat{c}_{i,\sigma l \alpha} \,
\end{aligned}
\end{equation}

Since the relatively large thickness makes big difference between intra-layer and inter-layer Coulomb interactions, thus it is necessary to employ a layer-dependent Coulomb interaction, denoted as:
\begin{equation}
\begin{aligned}
V_{ll}(\mathbf{q}) &= \frac{e^2}{2 \Omega_0 \epsilon_r \epsilon_0 \sqrt{q^2 + \kappa^2}} \\
V_{ll'}(\mathbf{q}) &= \frac{e^2}{2 \Omega_0 \epsilon_r \epsilon_0 q} e^{-q |l - l'| d_0}, \quad l \neq l'
\end{aligned}
\end{equation}
where the $\Omega_0$ refers to the area of moir\'e cell, and the inverse screening length $\kappa=0.025$ $\mathring{\mathrm{A}}$. Since the two valleys are well separated in real space, the inter-valley interaction can be neglected compared to the intra-valley interaction, which can be written as:
\begin{equation}
\hat{V}^{\mathrm{intra}}
= \frac{1}{2N_s} 
\sum_{\alpha \alpha'} \sum_{\mu \mu'} \sum_{\sigma \sigma'} 
\sum_{\mathbf{k} \mathbf{k}' \mathbf{q}} V_{ll'}(\mathbf{q})
\hat{c}^\dagger_{\sigma \mu l \alpha}(\mathbf{k}+\mathbf{q})
\hat{c}^\dagger_{\sigma' \mu' l' \alpha'}(\mathbf{k}'-\mathbf{q})
\hat{c}_{\sigma' \mu' l' \alpha'}(\mathbf{k}') 
\hat{c}_{\sigma \mu l \alpha}(\mathbf{k})
\end{equation}
where $N_s$ is the total number of the superlattice’s sites. Since our primary interest lies in the low-energy properties of the RMG–hBN system, the Coulomb interaction can be projected onto the band basis. The transformation of the
electron annihilation operators from the original basis to the band basis reads:
\begin{equation}
\hat{c}_{\sigma \mu l \alpha}(\mathbf{k}) \equiv \hat{c}_{\sigma \mu l \alpha \mathbf{G}}(\tilde{\mathbf{k}}) = \sum_{n} C_{\mu \alpha l \mathbf{G},n}(\tilde{\mathbf{k}}) \hat{c}_{\sigma \mu ,n}(\tilde{\mathbf{k}})
\end{equation}
where $C_{\mu \alpha l \mathbf{G},n}(\tilde{\mathbf{k}})$ is the expansion coefficient in the $n$-th Bloch eigenstate at $\mathbf{\tilde{k}}$ of valley $\mu$:

\begin{equation}
|\sigma \mu, n ; \tilde{\mathbf{k}} \rangle
= \sum_{\alpha \mathbf{G}} C_{\mu \alpha l \mathbf{G},n}(\tilde{\mathbf{k}}) | \sigma, \mu, l, \alpha, \mathbf{G}; \tilde{\mathbf{k}} \rangle
\end{equation}

We note that the non-interacting Bloch functions are spin degenerate due to the separate spin rotational symmetry
($SU (2) \otimes  SU (2)$ symmetry) of each valley. This allows us to express the intra-valley Coulomb interaction in the band
basis:
\begin{equation}
\begin{split}
\hat{V}^{\mathrm{intra}} = & \frac{1}{2N_s} \sum_{\tilde{\mathbf{k}} \tilde{\mathbf{k}}' \tilde{\mathbf{q}}}
\sum_{\substack{\mu \mu' \\ \sigma \sigma' \\ ll'}} \sum_{\substack{n m \\ n' m'}} \left( \sum_{\mathbf{Q}} V_{ll'} (\mathbf{Q} + \tilde{\mathbf{q}}) \Omega^{\mu l,\mu' l'}_{nm,n'm'} (\tilde{\mathbf{k}}, \tilde{\mathbf{k}}', \tilde{\mathbf{q}}, \mathbf{Q})
\right) \\
& \times \hat{c}^\dagger_{\sigma \mu ,n} (\tilde{\mathbf{k}} + \tilde{\mathbf{q}}) \hat{c}^\dagger_{\sigma' \mu' ,n'}(\tilde{\mathbf{k}}'-\tilde{\mathbf{q}}) \hat{c}_{\sigma' \mu' m'}(\tilde{\mathbf{k}}') \hat{c}_{\sigma \mu m} (\tilde{\mathbf{k}})
\end{split}
\end{equation}
where the form factor $\Omega^{\mu l,\mu' l'}_{nm,n'm'}$ reads:
\begin{equation}
\Omega^{\mu l, \mu'l'}_{nm,n'm'} (\tilde{\mathbf{k}}, \tilde{\mathbf{k}}', \tilde{\mathrm{q}}, \mathbf{Q})= \sum_{\alpha\alpha' \mathbf{G} \mathbf{G}'} C^*_{\mu l \alpha\mathbf{G+Q},n}(\tilde{\mathbf{k}}+\tilde{\mathbf{q}}) C^*_{\mu' l' \alpha'\mathbf{G'-Q},n'}(\tilde{\mathbf{k}}-\tilde{\mathbf{q}}) C_{\mu'l'\alpha'\mathbf{G}',m'}(\tilde{\mathbf{k}}') C_{\mu l \alpha \mathbf{G} ,m }(\tilde{\mathbf{k}})
\end{equation}

We apply the Hartree-Fock (HF) approximation to the Coulomb interaction, decomposing the two-particle Hamiltonian into the Hartree and Fock terms. The Hartree term is given by:
\begin{equation}
\begin{split}
\hat{V}_H^{\text{intra}} &= \frac{1}{2N_s} \sum_{\widetilde{\mathbf{k}}\widetilde{\mathbf{k}}'} \sum_{\substack{\mu\mu' \\ \sigma\sigma' \\ ll'}} \sum_{\substack{nm \\ n'm'}} \left( \sum_{\mathbf{Q}} V_{ll'}(\mathbf{Q}) \Omega_{nm,n'm'}^{\mu l,\mu' l'}(\widetilde{\mathbf{k}},\widetilde{\mathbf{k}}', 0, \mathbf{Q}) \right) \\
&\quad \times \left( \langle \hat{c}_{\sigma\mu,n}^{\dagger}(\widetilde{\mathbf{k}}) \hat{c}_{\sigma\mu,m}(\widetilde{\mathbf{k}}) \rangle \hat{c}_{\sigma'\mu',n'}^{\dagger}(\widetilde{\mathbf{k}}') \hat{c}_{\sigma'\mu',m'}(\widetilde{\mathbf{k}}') + \langle \hat{c}_{\sigma'\mu',n'}^{\dagger}(\widetilde{\mathbf{k}}') \hat{c}_{\sigma'\mu',m'}(\widetilde{\mathbf{k}}') \rangle \hat{c}_{\sigma\mu,n}^{\dagger}(\widetilde{\mathbf{k}}) \hat{c}_{\sigma\mu,m}(\widetilde{\mathbf{k}}) \right)
\end{split}
\end{equation}

and the Fock term reads:
\begin{equation}
\begin{split}
\hat{V}_F^{\text{intra}} &= -\frac{1}{2N_s} \sum_{\widetilde{\mathbf{k}}\widetilde{\mathbf{k}}'} \sum_{\substack{\mu\mu' \\ \sigma \\ ll'}} \sum_{\substack{nm \\ n'm'}} \left( \sum_{\mathbf{Q}} V_{ll'}(\widetilde{\mathbf{k}}' - \widetilde{\mathbf{k}} + \mathbf{Q}) \Omega_{nm,n'm'}^{\mu l,\mu' l'}(\widetilde{\mathbf{k}},\widetilde{\mathbf{k}}', \widetilde{\mathbf{k}}' - \widetilde{\mathbf{k}}, \mathbf{Q}) \right) \\
&\quad \times \left( \langle \hat{c}_{\sigma\mu,n}^{\dagger}(\widetilde{\mathbf{k}}') \hat{c}_{\sigma\mu',m'}(\widetilde{\mathbf{k}}') \rangle \hat{c}_{\sigma\mu',n'}^{\dagger}(\widetilde{\mathbf{k}}) \hat{c}_{\sigma\mu,m}(\widetilde{\mathbf{k}}) + \langle \hat{c}_{\sigma\mu',n'}^{\dagger}(\widetilde{\mathbf{k}}) \hat{c}_{\sigma\mu,m}(\widetilde{\mathbf{k}}) \rangle \hat{c}_{\sigma\mu,n}^{\dagger}(\widetilde{\mathbf{k}}') \hat{c}_{\sigma\mu',m'}(\widetilde{\mathbf{k}}') \right)
\end{split}
\end{equation}

After progressively integrating out the high-energy degrees of freedom via renormalization group \cite{vafek2020renormalization,guo2024fractional} from the high-energy cutoff $E_C \sim 2$ eV down to the low-energy cutoff $E_C^* \sim 0.15$ eV, we obtain a renormalized low-energy Hamiltonian that includes the contributions of Coulomb interactions from the remote-band electrons to the low-energy ones, together with a self-consistently screened $D$ field \cite{jang2023chirality,guo2024fractional}. To handle this model within the low-energy window, we keep three valence bands and three conduction bands per spin per valley ($n_{\text{cut}} = 3$) and project the electron-electron interactions onto the non-interacting wavefunctions of the renormalized low-energy effective model. It should be emphasized that within the low-energy window $E_C^*$, the long-range Coulomb interactions can no longer be treated perturbatively. Subsequently, we perform unrestricted self-consistent HF calculations within this low-energy window, assuming 32 possible initial symmetry-breaking states characterized by the order parameters $s_{0,z} \tau_a \sigma_b$ (with $a,b = 0, x, y, z$). A $18 \times 18$ $k$-mesh and 81 plane-wave components are used in the HF calculations.

\section{More phase diagrams of HF results}\label{phase}

\subsection*{Phase diagrams of RMG-hBN}
\begin{figure}
    \includegraphics[width=1\linewidth]{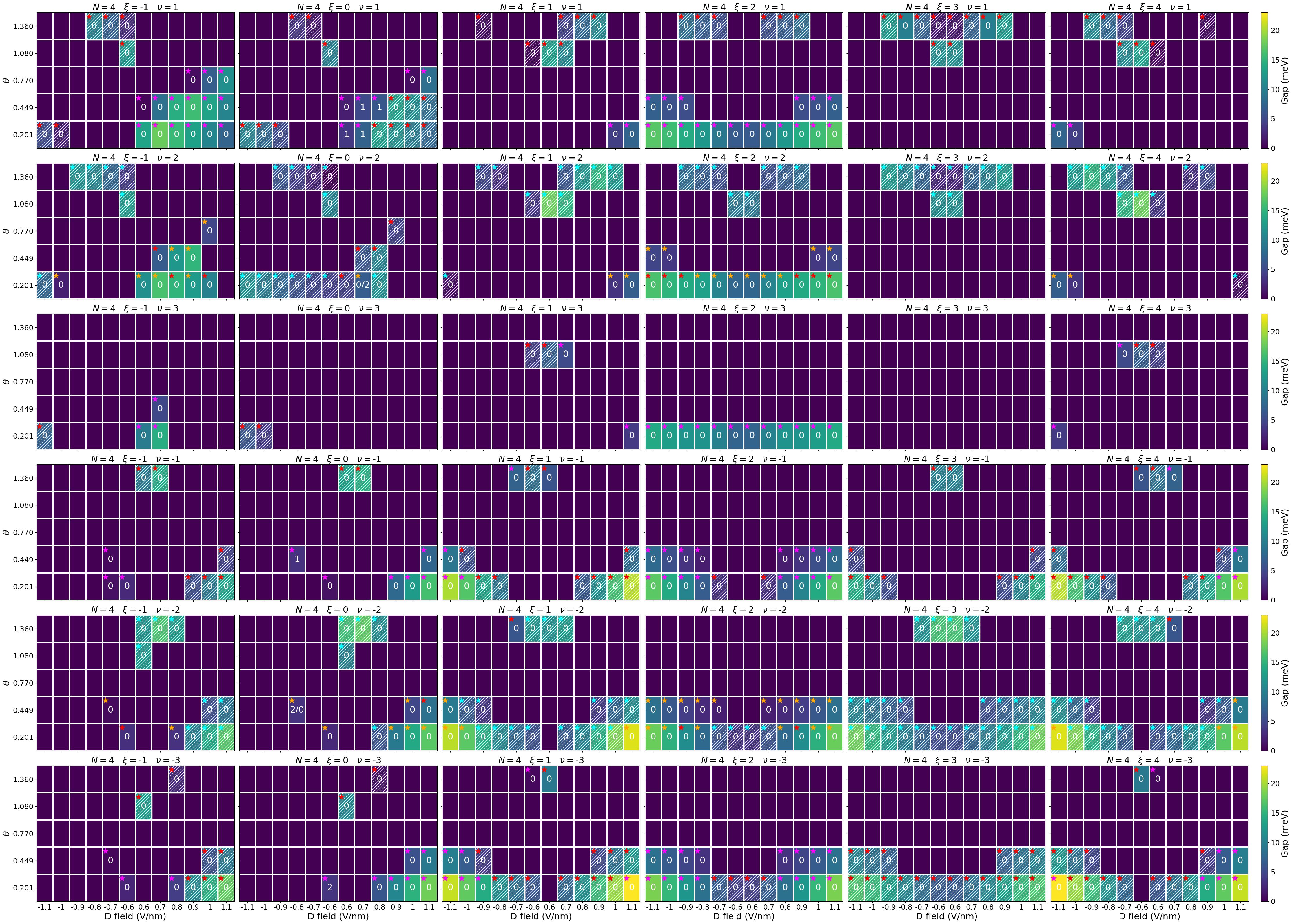}
    \caption{Phase diagram of the R4G-hBN system at different twist angles $\theta$, filling factor $\nu$ and hBN alignment configurations $\xi$. Stars of different colors — red, orange, magenta, and cyan — represent SP, VP, SVP, and UP, respectively. White hatching denotes the IVC state.} 
    \label{phase_start}
\end{figure}

The ground states of the RMG-hBN system are determined by self-consistently solving the HF equations, as shown in Fig. S\ref{phase_start} to S\ref{phase_end} for $N=4,\dots,9$. In our presented phase diagrams, the color-coded background represents the global band gap of the HF band structure, where metallic regions are denoted in deep purple. To account for numerical precision and potential quasi-degeneracies, we adopt a threshold of $0.01$ meV; states with energy differences below this value are considered degenerate, and their corresponding Chern numbers are separated by a slash (e.g., $1/0$) in the phase diagrams. Moreover, for a few parameter combinations, even though the overall gap is finite, the conduction band occupied by the doped electrons overlaps with the valence band, making it impossible to compute the Chern number. In such cases, we leave the Chern number unmarked in the phase diagram. An example is the case with $N=8$, $\xi=-1$, $\nu=3$, $\theta=0.201^\circ$, and $D=0.6$ V/nm. 

As explained in the main text, correlated insulating states are favored by small twist angles but are most abundant at intermediate layer number, reflecting the balance between bandwidth suppression and layer delocalization of the wavefunctions of active carriers, as also shown by Fig.~S\ref{rho_all_case} for $D>0, \nu<0$; $D<0, \nu>0$; and $D<0, \nu<0$. By Fig.~S\ref{xi_res}, the alignment-resolved results are presented.

\begin{figure}
    \includegraphics[width=1\linewidth]{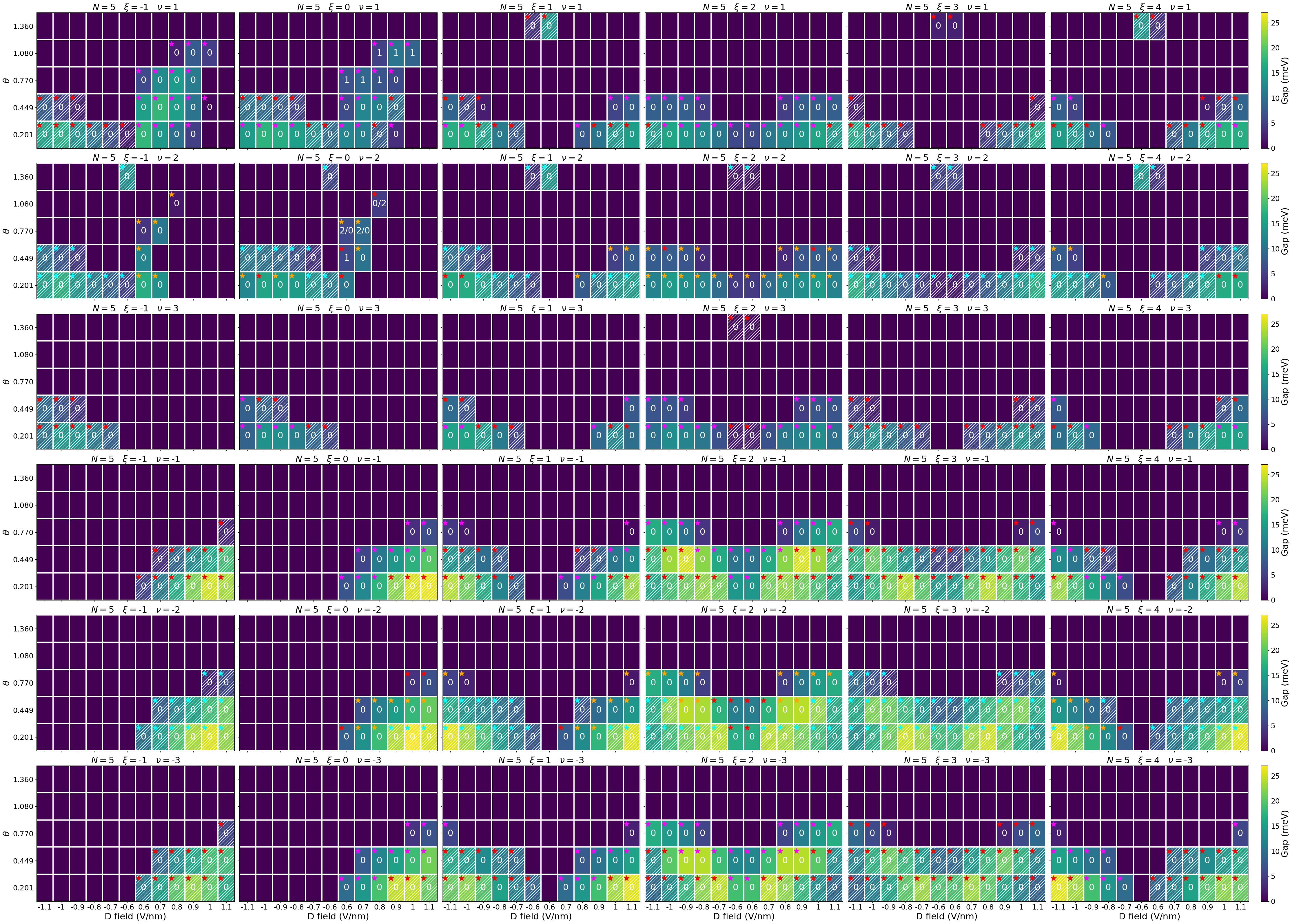}
    \caption{Phase diagram of the R5G-hBN system at different twist angles $\theta$, filling factor $\nu$ and hBN alignment configurations $\xi$. Stars of different colors — red, orange, magenta, and cyan — represent SP, VP, SVP, and UP, respectively. White hatching denotes the IVC state.}
\end{figure}

\begin{figure}
    \includegraphics[width=1\linewidth]{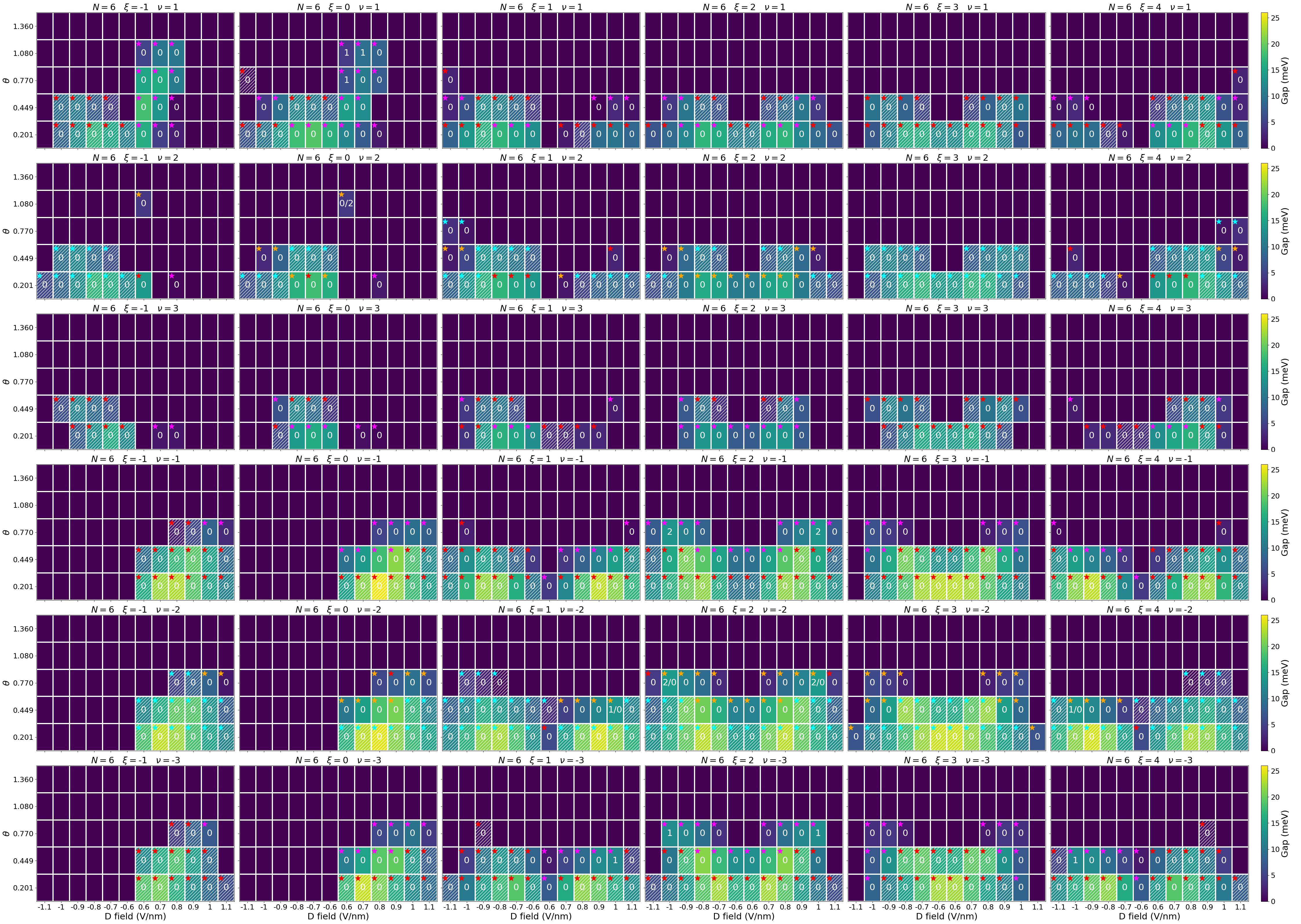}
    \caption{Phase diagram of the R6G–-BN system at different twist angles $\theta$, filling factor $\nu$ and hBN alignment configurations $\xi$. Stars of different colors — red, orange, magenta, and cyan — represent SP, VP, SVP, and UP, respectively. White hatching denotes the IVC state.}        
\end{figure}

\begin{figure}
    \includegraphics[width=1\linewidth]{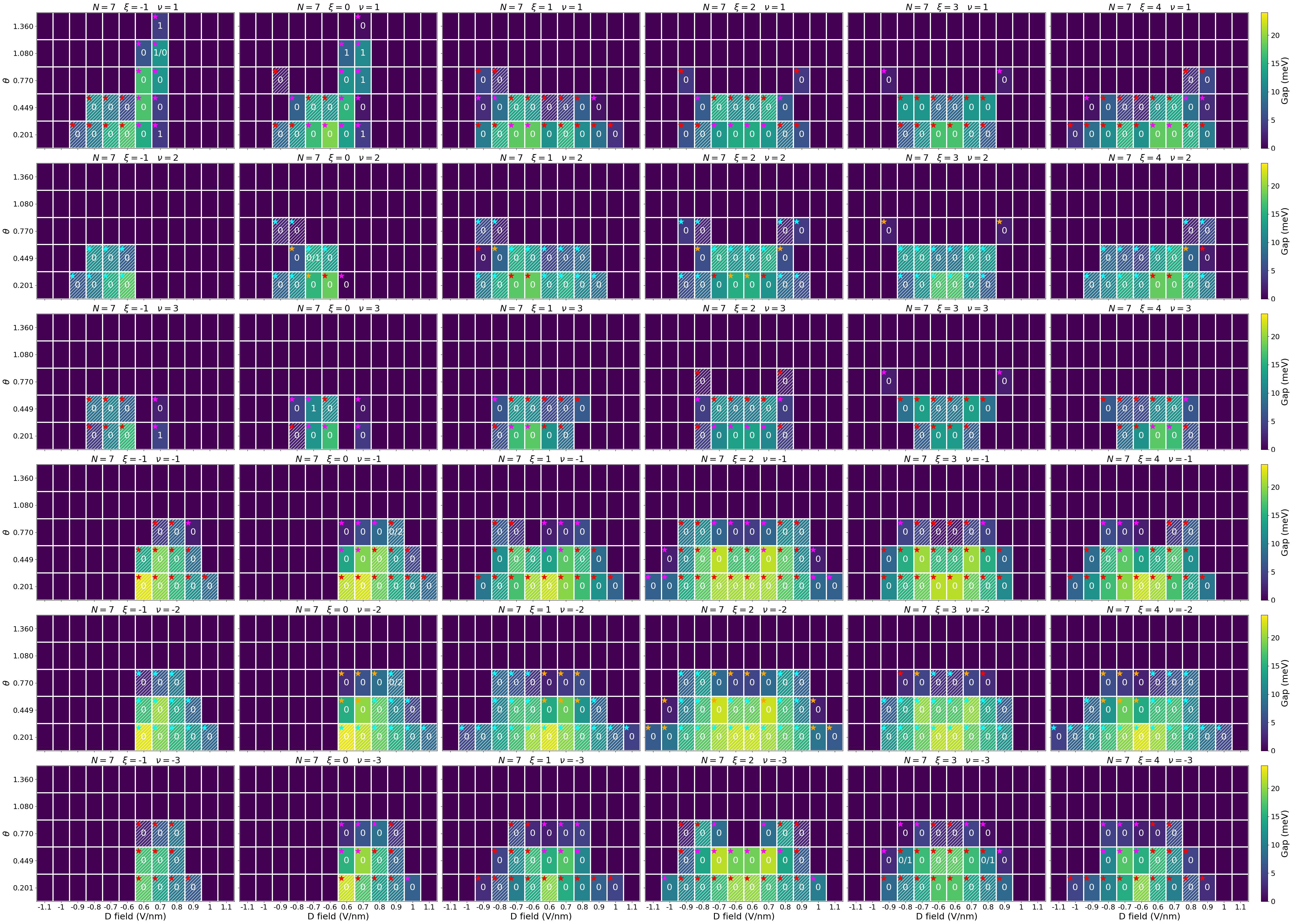}
    \caption{Phase diagram of the R7G-hBN system at different twist angles $\theta$, filling factor $\nu$ and hBN alignment configurations $\xi$. Stars of different colors — red, orange, magenta, and cyan — represent SP, VP, SVP, and UP, respectively. White hatching denotes the IVC state.}        
\end{figure}

\begin{figure}
    \includegraphics[width=1\linewidth]{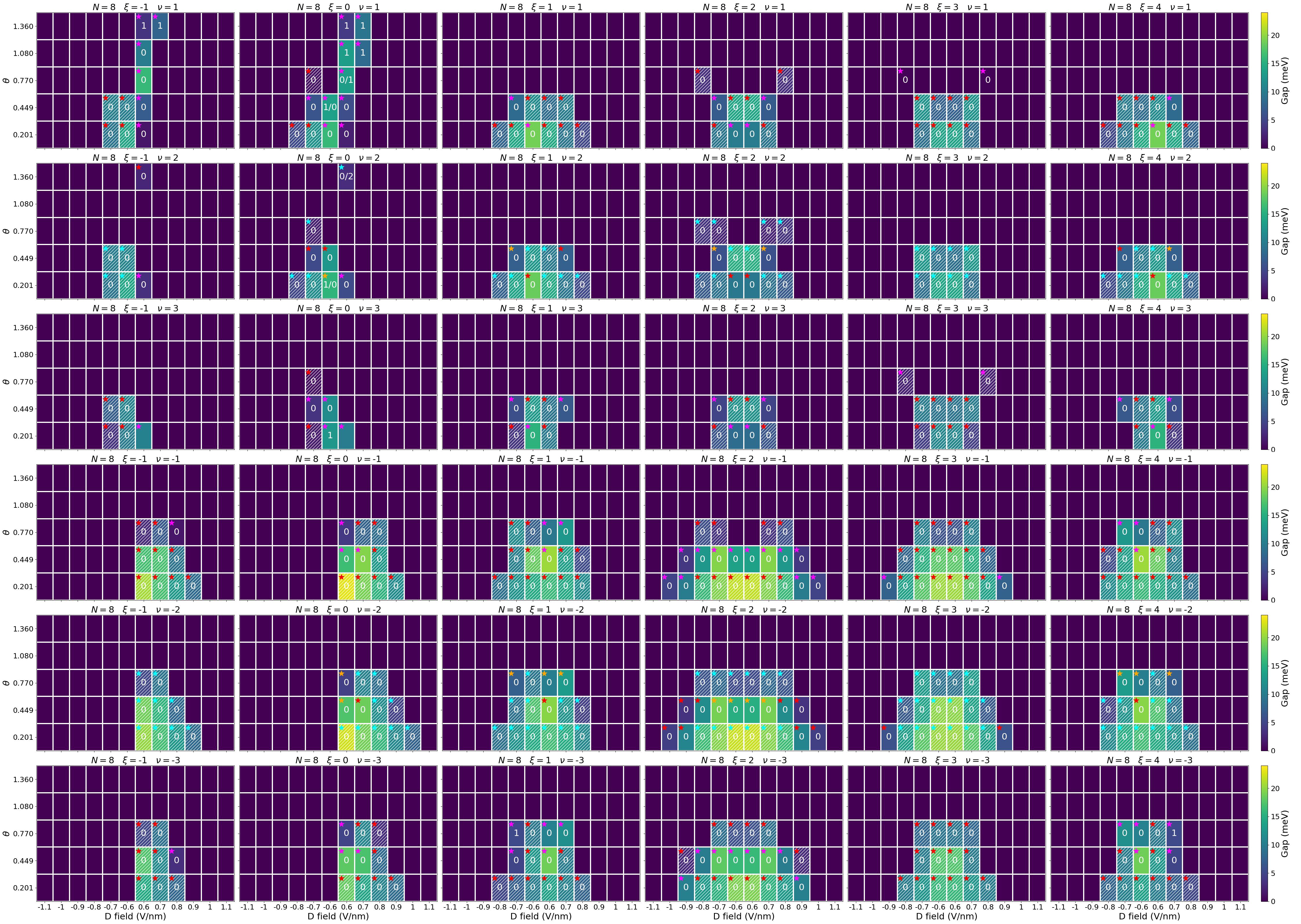}
    \caption{Phase diagram of the R8G-hBN system at different twist angles $\theta$, filling factor $\nu$ and hBN alignment configurations $\xi$. Stars of different colors — red, orange, magenta, and cyan — represent SP, VP, SVP, and UP, respectively. White hatching denotes the IVC state.}        
\end{figure}

\begin{figure}
    \includegraphics[width=1\linewidth]{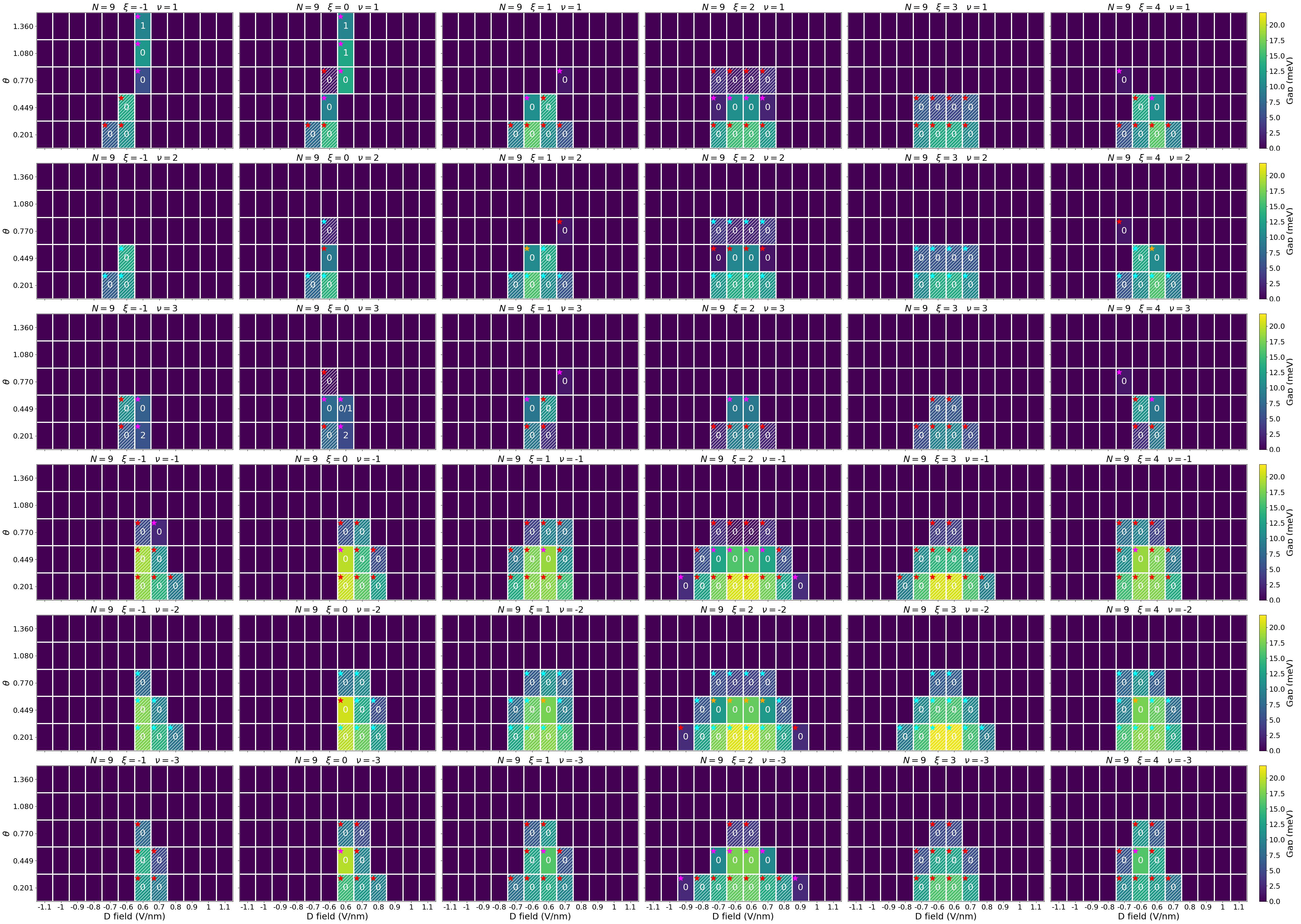}
    \caption{Phase diagram of the R9G–hBN system at different twist angles $\theta$, filling factor $\nu$ and hBN alignment configurations $\xi$. Stars of different colors — red, orange, magenta, and cyan — represent SP, VP, SVP, and UP, respectively. White hatching denotes the IVC state.}
    \label{phase_end}
\end{figure}

\begin{figure}
    \includegraphics[width=0.8\linewidth]{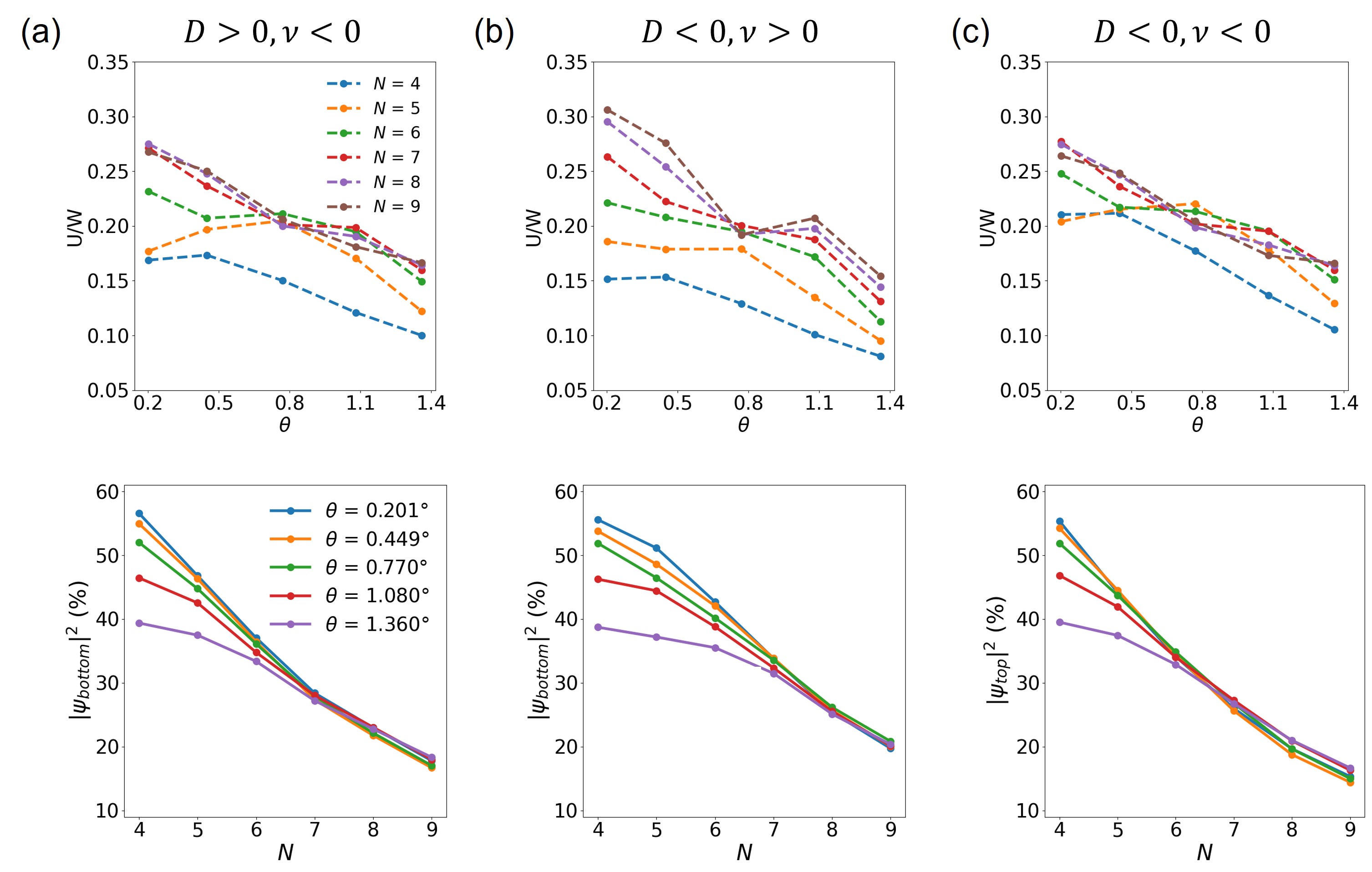}
    \caption{This figure shows two quantities. The upper panels show the $\theta$-dependence of the interaction strength as a function of $N$ (similar to Fig.~2(b) in the main text). The lower panels show the contribution from the top/bottom graphene layer to the active frontier bands (similar to Fig.~2(c)). Panels (a), (b), and (c) correspond to the three cases: $D>0, \nu<0$; $D<0, \nu>0$; and $D<0, \nu<0$, respectively.}
    \label{rho_all_case}
\end{figure}

\begin{figure}
    \includegraphics[width=1\linewidth]{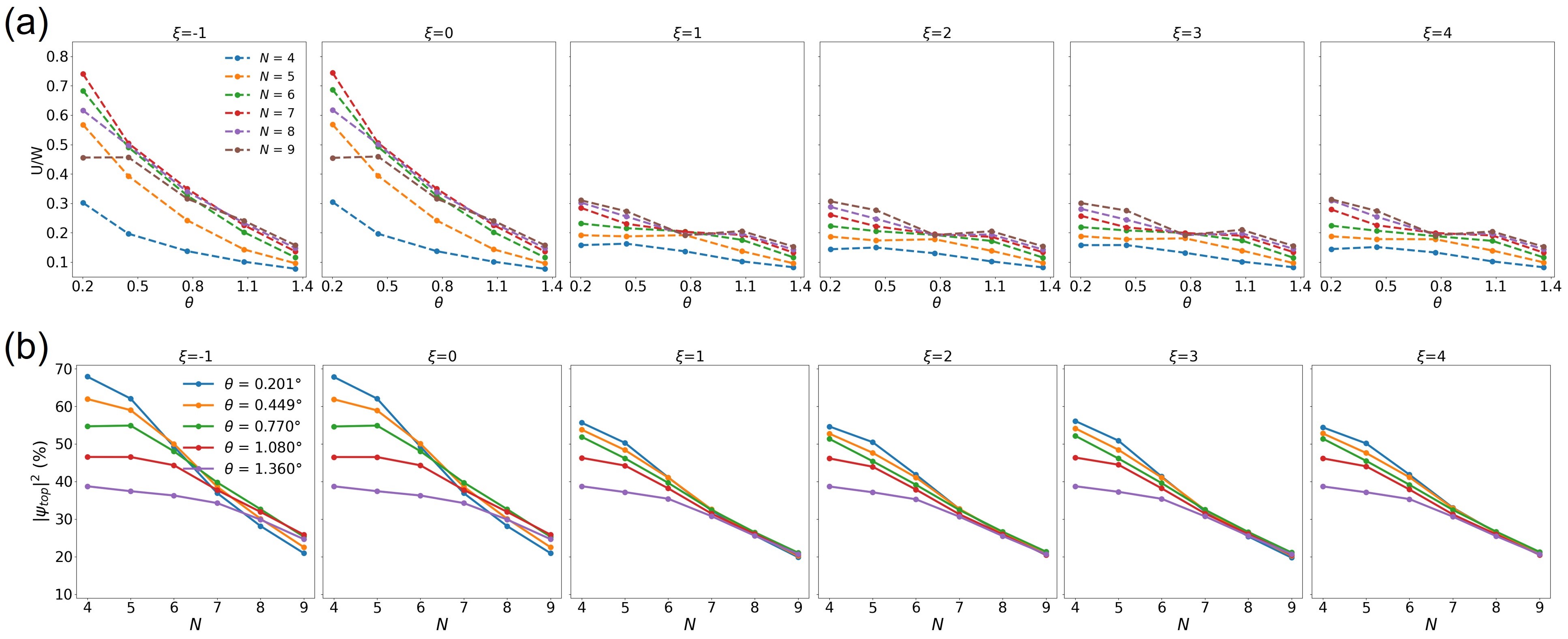}
    \caption{(a) and (b) present the $\xi$-resolved version corresponding to Fig.~2(b) and (c) in the main text, respectively: (a) interaction strength and (b) the contribution of top-layer graphene to the active frontier conduction band in the non-interacting limit.}
    \label{xi_res}
\end{figure}

\begin{figure}
    \includegraphics[width=0.5\linewidth]{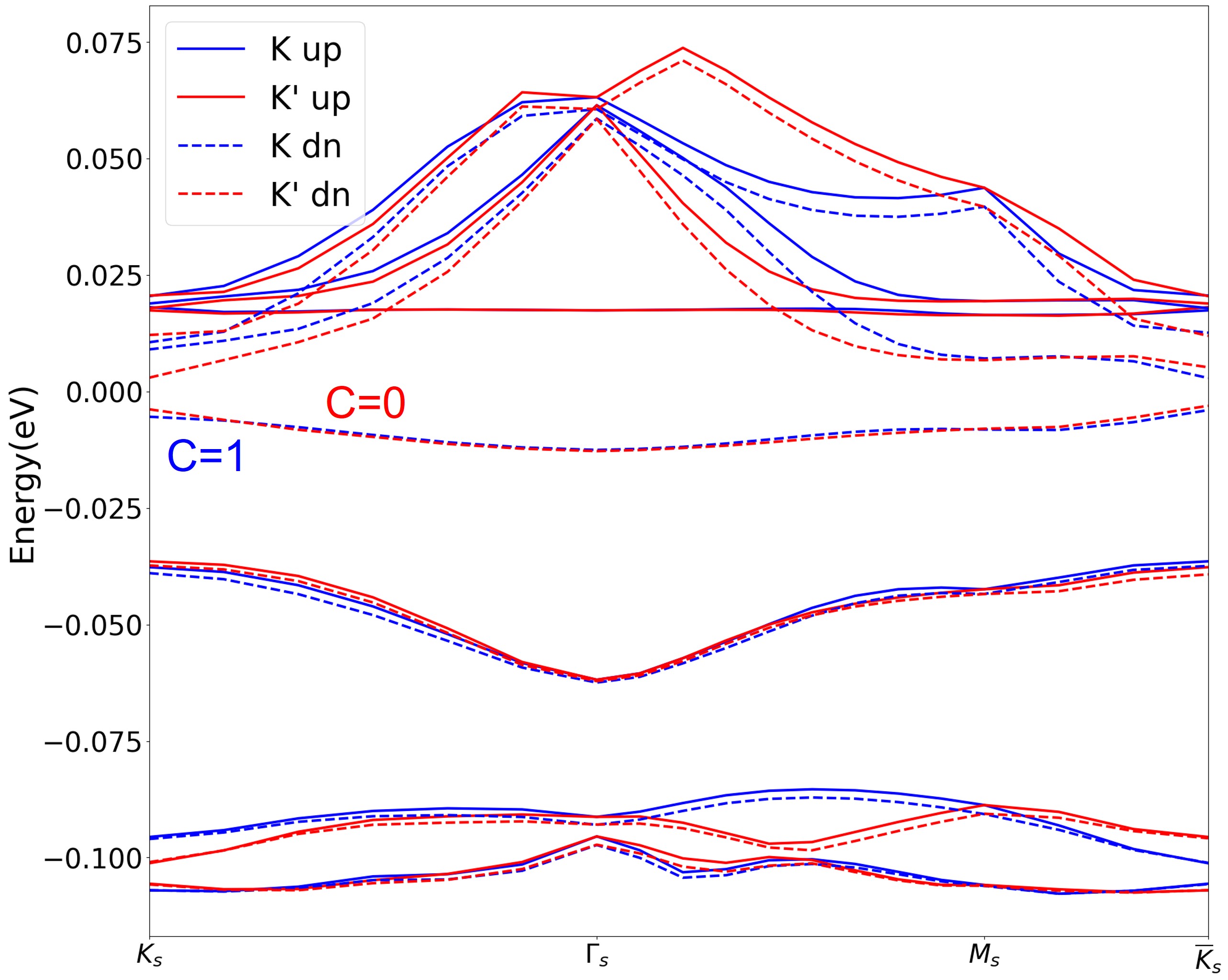}
    \caption{Spin-valley resolved HF band structure of R5G-hBN with $\theta = 0.449^\circ$, $D = 0.6$ V/nm, and $\nu = 2$, in which the two doped electrons result in a $C=1$ state.}
    \label{sep_chern}
\end{figure}

\begin{figure}
    \includegraphics[width=0.8\linewidth]{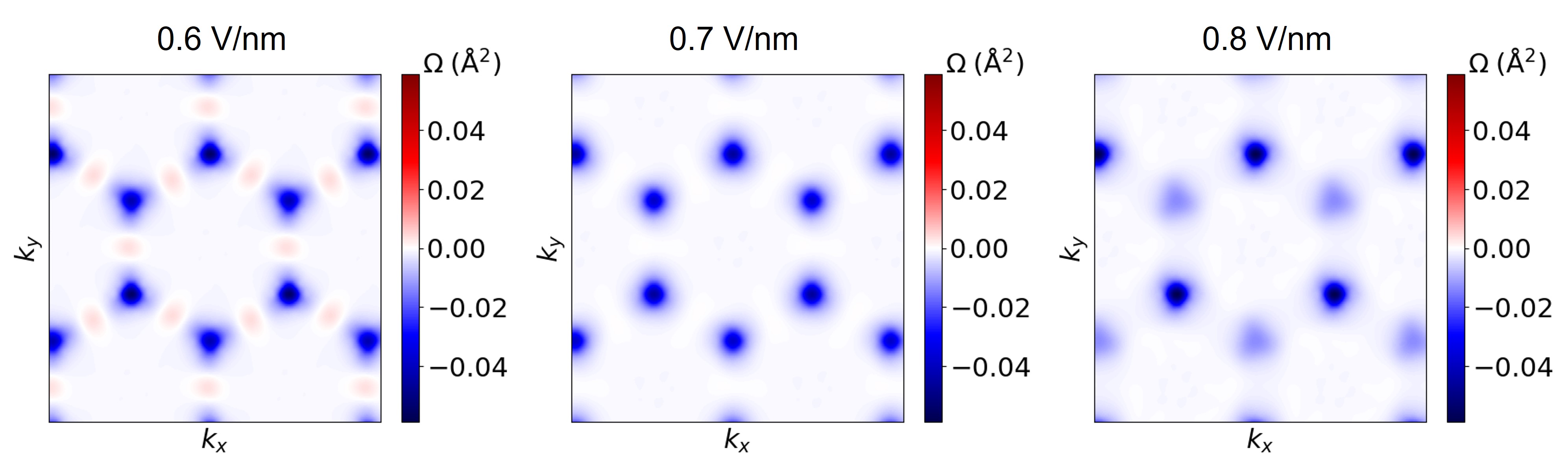}
    \caption{For parameters $\theta = 0.770^\circ$, $\nu = 1$, and $\xi = 0$, the R5G-hBN system exhibits $C = 1$ in the range $D = 0.6$--$0.8$ V/nm. The figure shows the Berry curvature distribution of the first conduction band in momentum space as a function of $D$.}
    \label{bcp_D}
\end{figure}

\begin{figure}
    \includegraphics[width=0.8\linewidth]{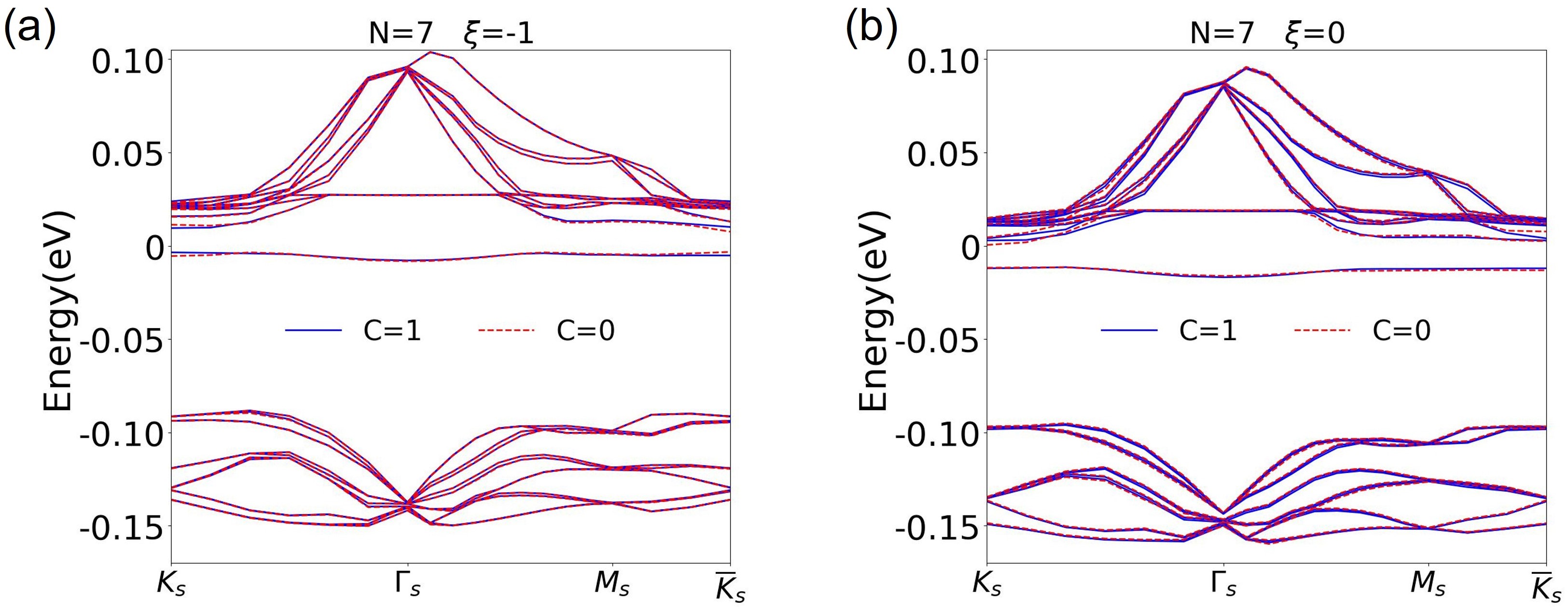}
    \caption{Panels (a) and (b) show the electronic structures for $\xi = -1$ and $0$, respectively, at $N = 7$, $\theta = 1.080^\circ$, $\nu = 1$, and $D = 0.68$~V/nm. For $\xi = -1$ ($\xi = 0$), the ground state is $C = 0$ ($C = 1$), and the opposite Chern state is metastable. As in the case of $N = 5$, the first conduction band, having nearly identical dispersions, exhibit completely different topological properties.}
    \label{N7band}
\end{figure}

\subsection*{Multiple flavor states and inter-valley coherent state}

To classify the symmetry-breaking patterns in the spin and valley degrees of freedom, we introduce the generalized order parameter defined as $\hat{O}_{abc} = s_a \otimes \tau_b \otimes I_{N \times N} \otimes \sigma_c$, where $s_a$, $\tau_b$, and $\sigma_c$ are the Pauli matrices acting on the spin, valley, and sublattice subspaces, respectively. The specific spin and valley configurations: Spin-Polarized (SP), Valley-Polarized (VP), Spin-Valley Polarized (SVP), and Unpolarized (UP), which are identified by evaluating the expectation values of these operators. For instance, the states illustrated in Fig. 4(a) and (b) in the main text exhibit $\langle \hat{O}_{z00} \rangle = 1$ and $\langle \hat{O}_{0z0} \rangle = 1$, which rigorously characterizes them as SVP phases where electrons occupy a single spin-valley flavor (e.g., $K$ valley with spin-up). Our numerical results demonstrate that these states with distinct spin-valley polarization can be accessed by tuning $D$, $\nu$ and $\theta$.

A noteworthy feature within the non-valley-polarized regime is the emergence of the inter-valley coherent (IVC) phase, represented by hatched regions in the phase diagrams. In this phase, the electronic wavefunctions from the $K$ and $K'$ valleys form a spontaneous superposition, characterized by non-vanishing expectation values of the off-diagonal operators $\langle \hat{O}_{0x(0,x,y)} \rangle$ or $\langle \hat{O}_{0y(0,x,y)} \rangle$. Physically, the IVC state locks the relative phase between the two valleys, thereby breaking the valley $U(1)$ symmetry. This symmetry-breaking mechanism is fundamentally distinct from and mutually exclusive with the valley polarization required for the CI state. Specifically, the phase coherence established in the IVC state precludes the well-defined valley-contrast necessary to host a non-zero Chern number in this system. This exclusion is clearly manifested in our phase diagrams, where the CI and IVC phases occupy distinct parameter regimes. While IVC states have been recently reported in other graphene moir\'e superlattices, our results provide a theoretical prediction for their manifestation and competition with topological phases in the RMG-hBN platform.

\subsection*{Specific examples for $N = 5$ and $7$}
This section summarizes several representative cases discussed in the main text.
\begin{itemize}
    \item Figure~S\ref{sep_chern}: a $\nu = 2$ insulating state with $C = 1$ is realized, where two isolated conduction bands carry Chern numbers $0$ and $1$, respectively, for example, for $N=5$, $\theta=0.449^\circ$, $\xi=0$ at $D=0.6$\;V/nm. \\
    \item Figure~S\ref{bcp_D}: For $N=5$, $\theta=0.77^\circ$, $\xi=0$ at $\nu=1$, the Berry curvature becomes concentrated at one of the moir\'e $K$ points with increasing $D$.\\
    \item Figure~S\ref{N7band}: The cases used for energy comparison in the main text ($N = 7$, $\theta = 1.080^\circ$, $\nu = 1$ under $D = 0.68$~V/nm), with $C = 1$ and $0$ for $\xi = 0$ and $-1$, show nearly identical electronic structures, indicating that interaction and kinetic energy differences can be safely neglected. 
\end{itemize}

\section{Evolution of real-space charge densities}\label{rho}
\begin{figure}
    \includegraphics[width=1\linewidth]{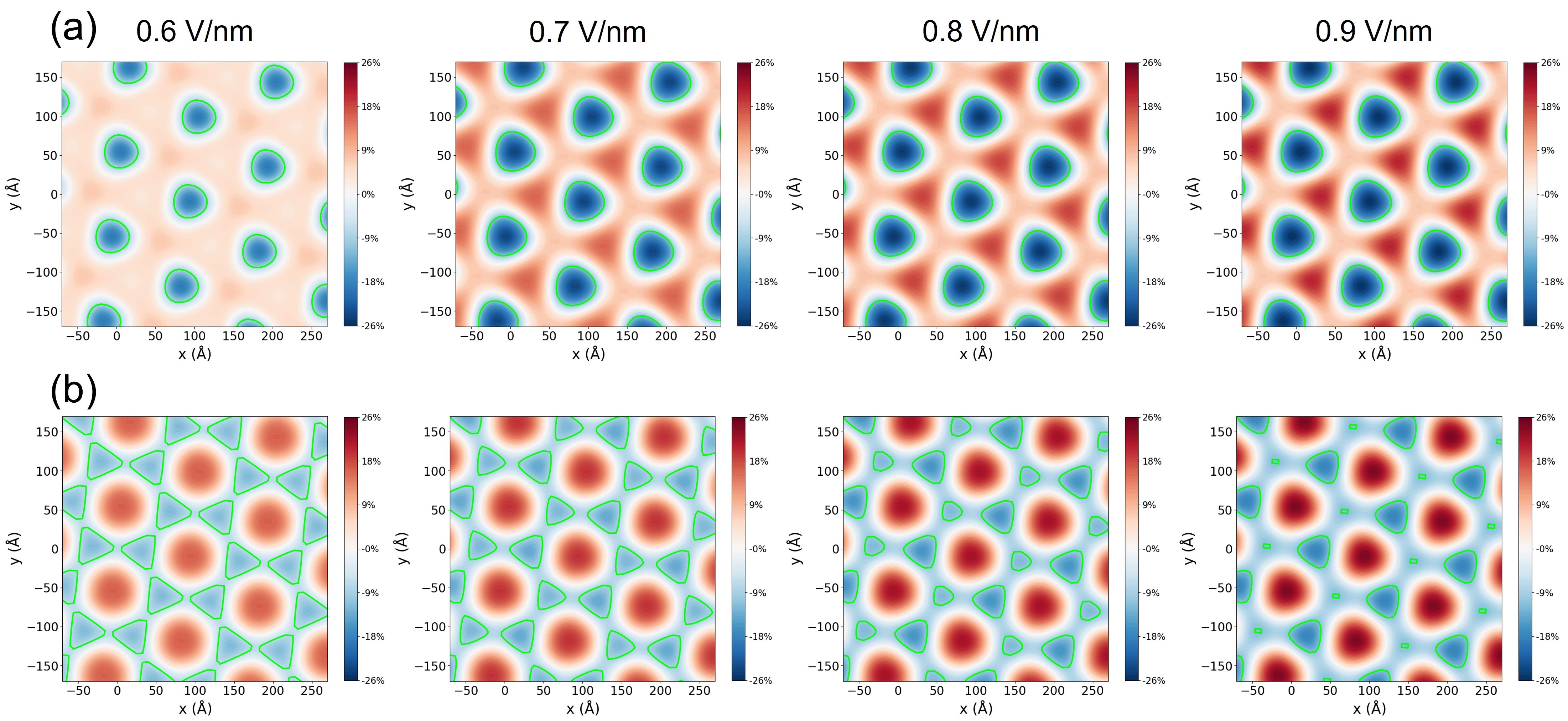}
    \caption{The real-space charge density of the valence band in the non-interacting limit is presented in (a) for $\xi=-1$ and in (b) for $\xi=0$. The used parameters are $N=5$, $\xi=-1$ and 0, $\theta=0.770^\circ$. The columns from left to right show the evolution of the density distribution under increasing displacement fields, specifically from $0.6$ to $0.9$ V/nm.}
    \label{rho_start}
\end{figure}

Figures S\ref{rho_start}-S\ref{rho_end} present the real-space charge densities calculated at the non-interacting, Hartree, and HF levels, for parameters $D = 0.6$--$0.9$ V/nm, $\xi = -1$ and $0$, $\theta=0.770^\circ$, $N = 5$, and $\nu = 1$. Note that the non-interacting first conduction band densities are not shown due to ubiquitous band entanglement between the first and higher conduction bands. By comparing the valence band charge densities across different levels of approximation, it is evident that the Hartree and Fock terms only slightly modulate the amplitude of the charge density without altering its overall spatial pattern. This suggests that the valence band distribution is predominantly determined by $\xi$ and $D$. Specifically, as the displacement field increases, the hexagonal patterns formed by the maxima in the $\xi=-1$ case (or the minima in the $\xi=0$ case) exhibit a characteristic distortion, where the intensities at adjacent corners show opposite trends: one increases while its neighbor decreases.

In contrast, the Hartree and Fock terms play a non-negligible role in the conduction band. The Hartree term introduces a charge compensation pattern for the doped electrons, which already manifests $\xi$-dependent localization and delocalization features. However, the Hartree-level density does not capture the mechanism of the $D$-driven topological phase transition from $C=1$ to $C=0$ in the $\xi=0$ alignment. It is the inclusion of the Fock exchange term that enables this transition. Although the $C=0$ state is raised in energy at the interband Hartree level, the intraband Fock exchange term lowers the energy of the $C=0$ state more significantly, thereby stabilizing it.

Consequently, we conclude that the moir\'e potential, which dictates the valence band charge density for different $\xi$, is fundamental for the alignment-dependent stability of CI states. The Hartree and Fock terms, acting upon this background charge distribution, further promote or suppress the emergence of these topological phases.

\begin{figure}
    \includegraphics[width=1\linewidth]{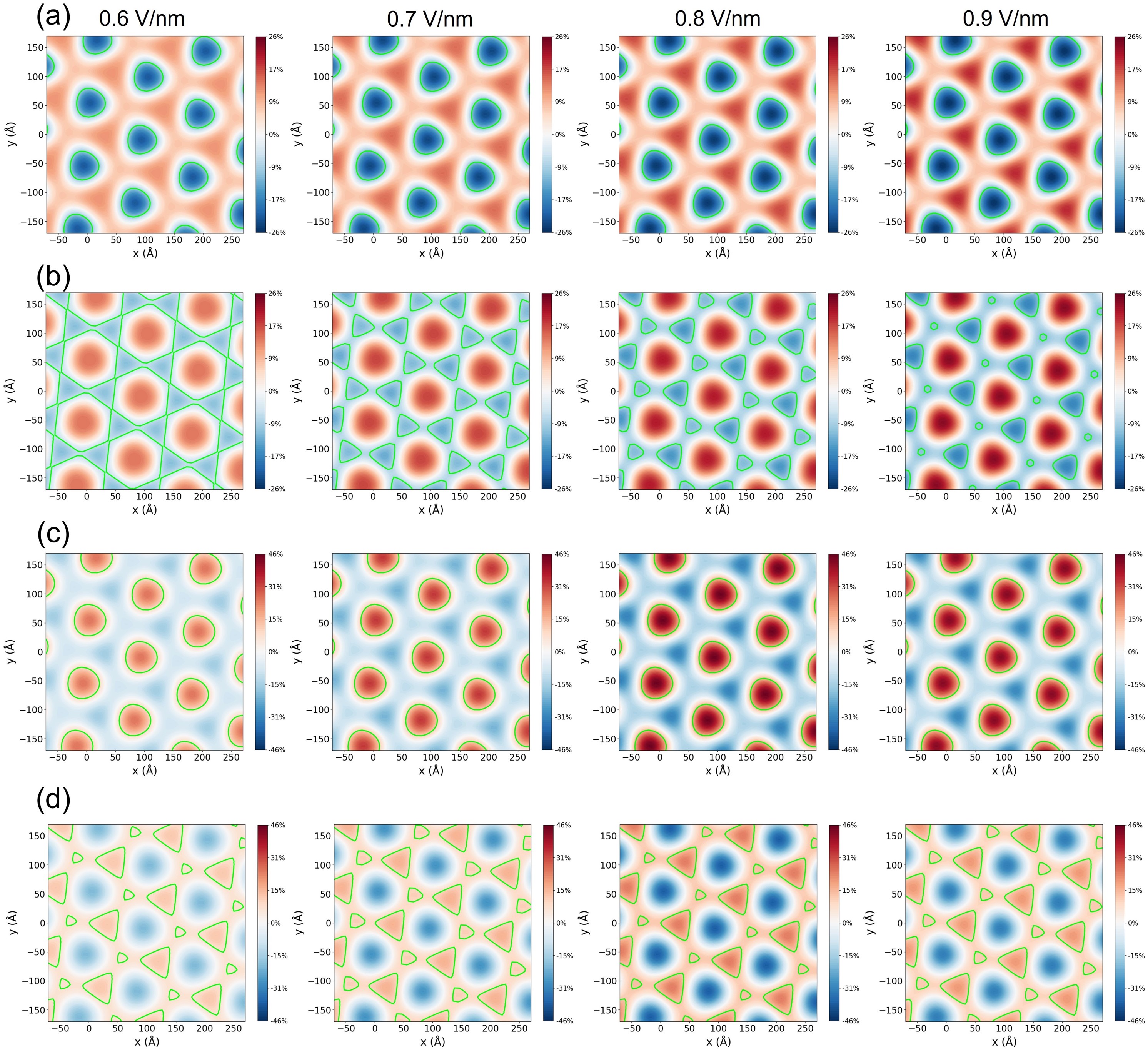}
    \caption{(a, b) Real-space charge densities of the valence band for $\xi=-1$ and $\xi=0$ alignments, respectively, calculated at the Hartree level. (c, d) Corresponding charge densities for the conduction band. The used parameters are $N=5$, $\xi=-1$ and 0, $\theta=0.770^\circ$. Each column represents a displacement field $D$ ranging from $0.6$ to $0.9$ V/nm.}        
\end{figure}

\begin{figure}
    \includegraphics[width=1\linewidth]{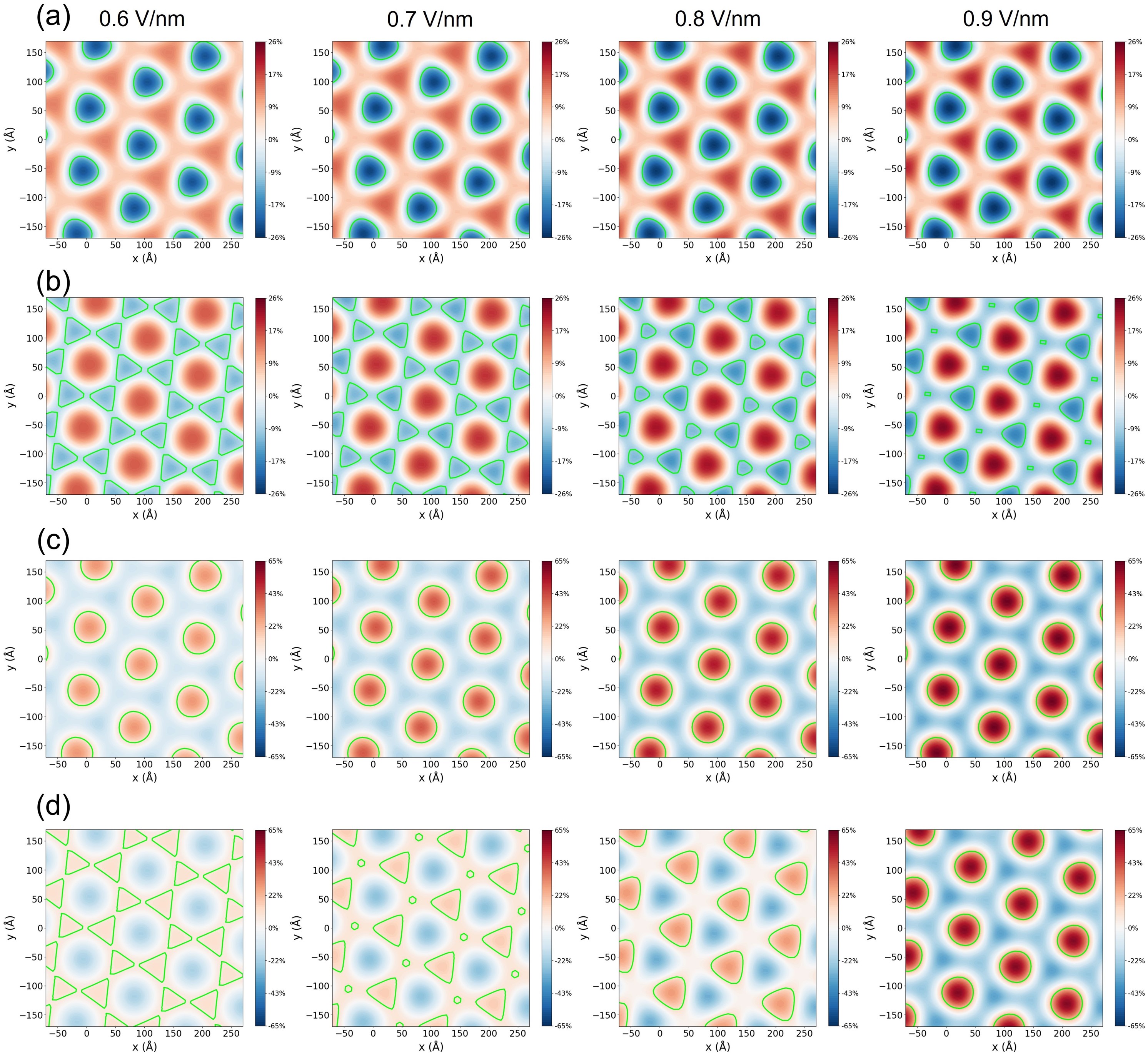}
    \caption{(a, b) Real-space charge densities of the valence band for $\xi=-1$ and $\xi=0$ alignments, respectively, calculated at the Hartree-Fock level. (c, d) Corresponding charge densities for the conduction band. The used parameters are $N=5$, $\xi=-1$ and 0, $\theta=0.770^\circ$. Each column represents a displacement field $D$ ranging from $0.6$ to $0.9$ V/nm.}
    \label{rho_end}
\end{figure}

\clearpage
\bibliographystyle{apsrev4-2}
\bibliography{ref}